\documentclass{aa}  
\usepackage{graphicx}
\usepackage{txfonts}

\begin{document}

\title{Asymmetry of the tidal tails of open star clusters 
  in direct $N$-body integrations in Milgrom-law dynamics}
\titlerunning{Discrete ML-dynamics of open star clusters}
\author{J. Pflamm-Altenburg
          \inst{1}}

   \institute{Helmholtz-Institut für Strahlen- und Kernphysik,
     Nussallee 14-16,
     D-53115 Bonn 
     \email{jpa@hiskp.uni-bonn.de}
   }

  \abstract
      {Numerical QUMOND-simulations
        of star clusters orbiting in a Galactic disk potential
        show that 
        the leading tidal arm of open star clusters
        contains tendentially more members than the trailing arm.
        However, these type of simulations are performed by
        solving the field-equations of QUMOND
        and already become non-practical for star cluster masses
        at around 5000\,$M_\odot$. Nearby star clusters have masses
        of 1000\,$M_\odot$ or $\approx$~1000 particles and less/fewer
        and can currently
        not be simulated reliably in field-theoretical formulations of MOND.}
      {The difference of the formation and evolution of tidal tails
       of open star clusters in the Newtonian and in the MONDian context
       is explored in the case of an equal-mass $n=400$  particle cluster ($M_\mathrm{tot}=200\,M_\sun$).}
      {In order to handle particle numbers below the QUMOND-limit
        the star cluster is simulated in Milgrom-law dynamics (MLD):
        Milgrom's law $g_\mathrm{N} = \mu(|a_\mathrm{M}|/a_0)a_\mathrm{M}$
        is postulated to be valid for discrete systems in vectorial form. 
        MLD shares the property with QUMOND that the acceleration
        of a particle outside any isolated mass concentration
        scales inversely with the distance. However, in MLD
        an internally
        Newtonian binary will follow a Newtonian rather than a MONDian
        path around the Galactic centre. In order to suppress the Newtonisation
        of compact subsystems in the star cluster
        the gravitational force is softened
        below particle distances of 0.001\,pc $\approx$ 206\,AU. Thus,
        MLD can
        only be considered as an approximation of a full MOND-theoretical
        description of discrete systems which are internally in the MOND
        regime.
        The MLD equations of motion are integrated by the
        standard Hermite scheme generally applied to Newtonian
        $N$-body systems, which is extended to solve for the accelerations
        and jerks associated with Milgrom's law.}
       {It is found that the tidal tails of a low-mass
        star cluster are populated
        asymmetrically in the MLD-treatment, very similar to the QUMOND
        simulations of the higher-mass star clusters.
        In the MLD-simulations the leading tail hosts
        up to twice as many members than the trailing arm
        and the low-mass open star cluster dissolves
        approximately 25\% faster than in the respective Newtonian
        case.
        Furthermore, the numerical simulations show that the
        Newtonian integrals of motion are not conserved
        in MLD.
        However, the case of an isolated binary
        in the deep MOND limit can be handled analytically.
        The velocity of the Newtonian centre of mass does not increase
        continuously but wobbles around the constantly moving MLD centre of
        mass.
   }
   {}


   \maketitle

   \section{Introduction}
   The rotation curves of disk galaxies are found to stay flat
   with increasing galactocentric radius. However,
   as the mass density decreases with increasing distance to the
   galactic centre the rotation curves were expected to fall
   \citep[e.g.][]{rubin1978a,bosma1981a}.
   This discrepancy has been attributed to the presence of
   unluminous (dark) matter. Candidates being part of the standard model of
   particle physics (e.g. brown dwarfs, white dwarfs and planetary objects
   have been continuously excluded \citep{bertone2018a}.
   Thus, it has been concluded that
   the hypothesised dark matter consists of particles beyond the standard
   model of particle physics. However, the direct search for dark matter
   particles is still negative. Even previously reported positive
   signals \citep{aprile2020a} could not be confirmed afterwards
   by the increased
   amount of measurement data \citep{aprile2022a}.

   Considered as one of the strongest direct empirical evidence for the
   existence of dark matter are the two colliding galaxy clusters, called
   Bullet cluster 1E~0657-56, where an offest between the peaks in the X-ray
   mass distribution of hot intra cluster gas  \citep{markevitch2002a}
   and the peaks in the
   mass distribution obtained by weak-lensing observations
   has been claimed to be detected \citep{clowe2006a}.
   However, it has been mentioned in \citet[][Sec. 4.3]{markevitch2002a}
   that the basic assumption of hydrostatic equilibrium required for the
   gas mass estimation can easily lead to an overestimate of the mass due to the high
   temperature during the collision.  
   Furthermore, the fulfillment of the basic requirements for the determination
   of the X-ray masses has never been put to the test in the case of the Bullet cluster.
   In order to assume hydrostatic equilibrium the sound crossing time
   through the cluster needs to be shorter than the age of the system
   \citep[e.g.][Sec.~2]{ettori2013a}. This might be the case for an
   isolated single galaxy cluster with an age in order of a Hubble time $\approx$~13~Gyr.
   
   The more massive main cluster in the Bullet cluster has a temperature of about 14~keV, the less massive
   subcluster a temperature of about 6~keV \citep{clowe2006a,markevitch2002a}.
   Taking a very conservative radius of $\approx 0.2$~Mpc of each cluster and the claimed peaks the sound crossing
   times for each subcomponents lie between 0.24~Gyr and 0.36~Gyr  and the sound crossing time of the
   entire system with an estimated radius of about 0.7~Mpc vary between 0.82~Gyr and 1.26~Gyr
   \citep[][Eq.~4]{ettori2013a}. \citet{barrena2002a} determined the age of the merger system to
   be about $0.15$~Gyr by tracing the orbits of the individual clusters back to the time point of impact.
   Thus, the requirement of hydrostatic equilibrium may be fulfilled by an isolated galaxy cluster but hardly by
   young merger systems like the Bullet cluster. 

   Furthermore, even if the derived X-ray gas distribution were true the standard argument that the hot plasma gas and
   the weak-lensing mass peaks are well separated from each other \citep[eg.][]{drees2019a} is not entirely true.
   Table~2 in \citet{clowe2006a} lists four mass peaks of the hot X-ray gas of similar total gas mass,
   two of them associated with the hot peaks which are offset from
   the two clusters. The two other peaks are located at the respective brightest cluster galaxies.

   As an alternative solution to the observed discrepancy between the
   observed distribution of matter in disk galaxies and their rotation
   curves \citet{milgrom1983a,milgrom1983b,milgrom1983c} proposed
   that the kinematical acceleration, $a$, is identical to the Newtonian
   gravitational acceleration, $g$, above a critical threshold
   $a_0 \approx 1.2\times 10^{-10}\,\rm m/s^2 = 3.8\,\rm pc/Myr^2$,
   \begin{equation}\label{eq_limit_newton}
     a = g\,,
   \end{equation}
   and is proportional to the square root of the Newtonian gravitational
   acceleration below $a_0$,
   \begin{equation}\label{eq_limit_mond}
     a = \sqrt{a_0Gg}\,,
   \end{equation}
   where $G = 0.0045\,\rm pc^3/M_\odot\,Myr^2$ is the gravitational constant.
   This concept is called Modified Newtonian Dynamics (MOND).
   The transition between the two regions is described by an interpolation
   function, $\mu(x)$,
   with properties
   \begin{equation}  \label{eq_trans_mu_fct}
     \mu(x) =\left\{
     \begin{array}{lcl}
       1 & , & x \gg 1\\
       x & , & x \ll 1\\
     \end{array}\right.
   \end{equation}
   and $\mu^\prime(x) > 0$, where the argument $x = a/a_0$ is the ratio of the
   absolute value of the kinematical acceleration and the threshold
   acceleration. Then, the relation between the kinematical and the gravitational
   acceleration is given by
   \begin{equation}\label{eq_ml}
     \mu(a/a_0)a = g\;,
   \end{equation}
   which is commonly referred to as Milgrom's law\, or in vectorial
   notation
   \begin{equation}\label{eq_mlv}
     \mu(|\mathbf{a}|/a_0)\mathbf{a} = \mathbf{g}\;.
   \end{equation}

   Shortly after the formulation of Eq.~(\ref{eq_ml})
   \citet{felten1984a} pointed out that the direct application of Eq.~(\ref{eq_ml})
   to the isolated two-body problem leads to a non-conservation of linear momentum.
   Such an unusual dynamical behavior is not surprising. As the gravitational
   acceleration, $\mathbf{g}$, on the right-hand side in
   Eq.~(\ref{eq_mlv}) is a conservative field,
   the kinematical acceleration, $\mathbf{a}$, on the left-hand side is generally
   not conservative.

   In order to ensure a conservative acceleration field \citet{bekenstein1984a}
   extended the classical Poisson equation for the Newtonian potential,
   $\Phi_\mathrm{N}$,
   \begin{equation}\label{eq_poisson}
     \Delta \Phi_\mathrm{N} = 4\pi G \rho\,,
   \end{equation}
   and formulated the AQUAL version of
   MOND,
   \begin{equation}\label{eq_aqual}
     \nabla\left( \mu\left(\frac{|\nabla\Phi_\mathrm{A}|}{a_0}\right)
     \nabla\Phi_\mathrm{A}\right)
     = 4\pi G \rho\,,
   \end{equation}
   where $\rho$ is the spatial mass density and $\Phi_\mathrm{A}$ is the
   potential of the kinematical acceleration field with the relation
   $\mathbf{a}=-\nabla\Phi_\mathrm{A}$. The AQUAL-Eq.~(\ref{eq_aqual})
   reduces to the vectorial form of Milgrom's law (Eq.~(\ref{eq_mlv}))
   only in cases of very high symmetry, e.g. in
   systems with a spherically symmetric mass distribution or in
   thin axis-symmetric disks.

   Due to the non-linearity  of Eq.~(\ref{eq_aqual})
   obtaining analytic 
   solutions  is much more difficult
   than in the simpler Poissonian case. \citet{milgrom2010a}
   formulated a quasi-linear version of MOND, where the differential part
   is identical to the Poissonian case but with a modified source term,
   \begin{equation}\label{eq_qumond}
     \Delta\Phi_\mathrm{Q} = \nabla\bullet
     \left[\nu\left(\frac{|\nabla\Phi_\mathrm{N}|}{a_0}\right)\nabla\Phi_\mathrm{N}\right]\,,
   \end{equation}
   where the Newtonian potential is given by the standard
   Poisson equation Eq.~(\ref{eq_poisson}).
   
   Thus, the QUMOND equation is linear in the QUMOND-potential,
   $\Phi_\mathrm{Q}$,
   but non-linear in the corresponding Newtonian potential. 
   Here, $\nu(y)$ is
   the transition function from the Newtonian to the MONDian regime
   with properties
   \begin{equation}  \label{eq_trans_nu_fct}
     \nu(y) =\left\{
     \begin{array}{lcl}
       1 & , & y \gg 1\\
       1/\sqrt{y} & , & y \ll 1\\
     \end{array}\right.
   \end{equation}
   in order to fulfill the observed boundary conditions
   in Eqs.~(\ref{eq_limit_newton},\ref{eq_limit_mond}).

   If the hidden mass problem is due to a change of the gravitational
   or the dynamical laws in the weak acceleration regime, then deviations
   from the Newtonian dynamics are expected on small scales as well, for example
   in Globular and open star clusters.
   Stellar tidal tails of star clusters are expected to form symmetrically
   in Newtonian dynamics \citep[eg.][]{kuepper2010a,pflamm-altenburg2023a}.
   Observations of the Galactic Globular cluster Pal~5 show asymmetries 
   between both tidal arms \citep{ibata2017a}. This has been attributed to
   a disruptive encounter with a dark matter sub halo \citep{erkal2017a}
   or a giant molecular cloud \citep{amorisco2016a}.
   In these cases, the observed asymmetry arises from
   a gap in previously symmetrically populated tidal tails. Contrary,
   in MONDian dynamics an asymmetry between both tidal tails
   arises naturally \citep{thomas2018a}.

   Asymmetries between leading and trailing tidal arms have also been found
   in four open star clusters (Hyades, Coma Berenices, Praesepe and
   NGC~752) in the Solar vicinity
   \citep{jerabkova2021a,boffin2022a,beccari2022a}.
   Dynamical interactions with a Galactic bar or spiral arms might
   cause local asymmetries in the tidal tails of open star clusters
   \citep[cf.]{bonaca2020a,pearson2017a}. In three cases (Praesepe,
   Coma Berenices and NGC~752) the degree of the observed asymmetry
   can be explained by the stochastic nature of the evaporation of single
   stars through both Lagrange points. 
   But in the case of the Hyades the random occurrence of the observed
   asymmetry would be a 6.7~$\sigma$ event \citep{pflamm-altenburg2023a}.
   Contrary, the asymmetric tidal tails of the Hyades are a very
   likely result in MONDian dynamics \citep{kroupa2022a}.   
   
   The dynamical simulation of the evolution of
   the tidal tails of the Globular cluster
   Pal~5 \citep{thomas2018a}
   and a massive open star cluster in the Galactic disk \citep{kroupa2022a}
   have been done by use of the
   Phantom of Ramses code {(\sc PoR}) \citep{lueghausen2015a}
   which is an extension of the Ramses code
   \citep{teyssier2002a} and allows the dynamical simulations 
   of systems containing gas and stars in the QUMOND-field formulation
   \citep{milgrom2010a}.

   Because MOND is formulated by the field theories
   AQUAL and QUMOND, numerical tools 
   require a smooth mass density distribution.
   Therefore, star clusters are treated to be collision-less systems
   and energy redistribution between individual particles are not considered.
   Particles can only leave the cluster if their effective energy in
   the co-rotating reference frame, where the star cluster is at rest, is large enough
   to pass through the Lagrange points and to overcome the tidal threshold.
   Stars having initially positive energy with respect to minimum energy required to
   escape from the cluster do not leave immediately the cluster but
   can be kept trapped by the star cluster up to a Hubble time \citep{fukushige2000a}.
   As energy redistribution between particles is not possible in collision-less $N$-body
   codes the upper region in the velocity distribution of stars is continuously depleted.
   However, the dynamics between the particles of discrete stellar systems repopulates
   the upper velocity region which leads to an enhanced evaporation. This special property
   of discrete stellar systems can only be modeled with a collisional code.
   The probability of escape from a star cluster is also effected by the orientation of the
   stellar orbit with the cluster \citep{read2006a,tiongco2016a} and it might be possible that the process of energy redistribution
   among stars is different in Newtonian and MONDian dynamics and leads already to an asymmetric
   evaporation process. This kind of dynamical issues can never be explored by a collision-less code. 
   Furthermore, the graininess of the mass distribution already leads to resolution limitation in
   the collision-less QUMOND modeling of a $5000~M_\odot$ star cluster \citep{kroupa2022a}.

   It is the aim of this work to construct equations of motion for a discrete
   stellar system in MONDian dynamics 
   and to study the formation and the evolution of tidal tails of low-mass star
   open clusters in this context. As equations of motion of a discrete stellar system
   are not available in the context of AQUAL and QUMOND the generalization of Milgrom's law
   is considered.
   In Sect.~\ref{sec_eqm} the equations of motion of a discrete stellar system
   are formulated.
   Section~\ref{sec_numerical_model} presents
   the numerical method to integrate the discrete equations of motion, which is
   an extension of the standard Hermite scheme used in direct Newtonian
   models. In Sect.~\ref{sec_num_tests} numerical tests of MLD are performed
   in order to explore the range of applicability and the limitation
   of this kind of MOND formulation.  
   The results of the numerical simulations are presented
   in Sect.~\ref{sec_simulations} where the formation and the evolution
   of stellar tidal tails of an open star cluster in the MONDian and the pure
   Newtonian case are compared with each other.

   \section{MLD equations of motion of a discrete stellar system}\label{sec_eqm}
   The direct application of Milgrom's law in order to calculate a MONDian acceleration
   from the Newtonian gravitational accelerations has been considered already
   in the past
   in a cosmological context \citep{nusser2002a,knebe2004a}. The gravitational
   acceleration was calculated using the Poisson equation where the smooth mass density distribution
   is obtained by a particle-mesh method.
   
   Here, this way is followed 
   and we postulate the validity of Milgrom's law in Eq.~(\ref{eq_mlv})
   to be valid for discrete systems.
   The equations of motion
   of an isolated, discrete system of $N$ gravitationally interacting
   particles are given by
   \begin{equation}\label{eq_mld}
     \mu\left(\frac{|\mathbf{a}_i|}{a_0}\right)\mathbf{a}_i
     =G \sum_{\substack{j=0\\j\neq i}}^{j=N} \frac{m_j}{\left|\mathbf{r}_j-\mathbf{r}_i\right|^3}\left(\mathbf{r}_j-\mathbf{r}_i\right)\,,
   \end{equation}
   where $m_j$ is the mass and $\mathbf{r}_j$ the position vector
   of the $j$-th particle, $\mathbf{r}_i$ the position vector
   and $\mathbf{a}_i$ the acceleration vector of the $i$-th particle.
   
   If a compact subsystem of $n$ particles is considered, e.g., a star cluster
   in a galaxy, the right hand
   side in Eq.~(\ref{eq_mld}) is split into two parts: i) the fully gravitationally
   self-interacting subsystem, ii) the embedding external gravitational mass, i.e.,
   all particles with index $i>n$, acting as an external perturbative Newtonian
   potential,
   $\Phi_\mathrm{ext}(\mathbf{r})$,
   \begin{equation}\label{eq_mld_sub}
     \mu\left(\frac{|\mathbf{a}_i|}{a_0}\right)\mathbf{a}_i
     =G \sum_{\substack{j=0\\j\neq i}}^{j=n} \frac{m_j}{\left|\mathbf{r}_j-\mathbf{r}_i\right|^3}\left(\mathbf{r}_j-\mathbf{r}_i\right)
     - \nabla \Phi_\mathrm{ext}(\mathbf{r}_i)\;.
   \end{equation}

   This $N$-body formulation of MOND also shows an external field effect (EFE)
   in the case where the internal acceleration is smaller than the external
   kinematical acceleration $a_\mathrm{ext}$. The scaling factor on the
   left-hand side in Eq.~(\ref{eq_mld_sub})
   is approximately a common constant factor $\mu(a_\mathrm{ext}/a_0)$ and
   the equation of motion are effectively those of a Newtonian system
   with a rescaled effective gravitational constant
   $G_\mathrm{EFE} = G /\mu(a_\mathrm{ext}/a_0)$.  
   For example the Pleiades open star cluster has a total stellar mass
   of about $740\,M_\odot$ and a half-mass radius of $r_\mathrm{h}=3.66\,\rm pc$
   \citep{pinfield1998a}. Treating the Pleiades as a Plummer sphere
   \citep{plummer1911a} 
   this corresponds to a Plummer parameter of $b=2.8\,\rm pc$.
   In a Plummer sphere the maximum internal acceleration is
   $a_\mathrm{max} = 2 M G / \sqrt{27} b^2$ at a radial distance of
   $r_\mathrm{max}=b/\sqrt{2}$ to the cluster centre. 
   For the Pleiades the maximum acceleration is then 0.16\,pc/Myr$^2\approx a_0/24$ and
   evolves as a quasi-Newtonian system with an increased Gravitational constant.

   As already mentioned in the Introduction the MONDian acceleration
   field is not
   conservative and the (classical) linear (Newtonian) momentum is not conserved
   \citep{felten1984a}, which means that
   it can not be derived as the gradient of a scalar potential.
   However, if the
   physical origin of MONDian effects
   is due to an interplay between the vacuum and the gravitating masses
   \citep{milgrom1999a}, 
   conservation of extensive quantities would be expected to exist
   for the total system
   (vacuum plus gravitating mass) rather than for an effectively
   non-isolated subsystem. Furthermore, \citet{felten1984a} restricted the discussion
   of momentum conservation to the isolated two-body problem in the deep MOND regime.
   Systems where direct tests of the conservation of linear momentum are
   accessible (surface of the earth and the solar system) have absolut acceleration
   well above the threshold acceleration, $a_0$, and are in the deep Newtonian regime.
   The vacuum only acts as a very weak perturber
   and its influence is basically negligible and no violation of the conservation
   of linear momentum will be detectable. But if the system has absolut
   accelerations below $a_0$ then the vacuum might have a strong influence on the
   dynamical properties of the gravitating system and the mass points
   can not be considered
   as an isolated system anymore.

   Furthermore, the non-existence of a scalar potential of the acceleration field
   in Eq.~(\ref{eq_mlv}) does not
   imply that Eq.~(\ref{eq_mld})
   is not equivalent to an Euler-Lagrange equation derivable from a Lagrangian.
   For instance, the equation of motion of
   a particle with mass $m$ and charge $q$ moving in an electric
   field $\vec{E}$ and non-conservative magnetic field $\vec{B}$,
   \begin{equation}
     m\ddot{\vec{q}} = q\left(\vec{E}+\dot{\vec{q}}\times\vec{B}\right)\,,
   \end{equation}
   can be derived from the Lagrangian
   \begin{equation}
     L = \frac{m}{2}\dot{\vec{q}}^2-q(\Phi-\dot{\vec{q}}\bullet\vec{A})\,,
   \end{equation}
   where $\vec{A}$ is the vector potential of the magnetic field
   $\vec{B}$.
   As a second example the time-independent equation of motion
   of the damped harmonic oscillator
   \begin{equation}\label{eq_eom_dho}
     m\ddot{q}+\gamma m\dot{q}+\omega^2q=0
   \end{equation}
   can be derived from the Lagrangian
   \begin{equation}
  L = e^{\gamma t}\left(\frac{m}{2}\dot{q}^2-\frac{\omega^2}{2}q^2\right)\,,
   \end{equation}
   even though the system loses continuously energy due to a non-conservative
   frictional force ($\gamma < 0$), reflected be the explicit time-dependency
   of the Lagrangian.

   Therefore, more theoretical work is required to explore if or to what extend
   Milgrom's law dynamics or related formulations can be expressed
   by a variational principle.

\section{Numerical algorithm}\label{sec_numerical_model}
As the numerical effort of integrating the
Newtonian equations of motion of a self-gravitating discrete system
scales to first order with the square of the particle number, these
equations are generally integrated by use of the Hermite scheme
\citep{aarseth2003a,makino1991a,kokubo1998a,hut1995a} to reduce
the number of function evaluations.
The Hermite method makes use of
the accelerations, $\vec{a}$,
calculated by direct summation over all pairwise interactions and their
time derivatives, $\vec{j}:=\dot{\vec{a}}$ (called jerks), because
the jerks can be obtained simultaneously during the calculation of the
accelerations.
In the following subsection the Hermite scheme, which
is a predictor corrector method,
is summarized.

\subsection{Hermite scheme}
Consider a particle at time $t$ with position
vector $\vec{r}$ and velocity vector $\vec{v}$. In order to advance
the particle in time with a time step of $dt$ the acceleration vector
$\vec{a}$ and its time derivative $\vec{j}=\dot{\vec{a}}$
 needs to be calculated at time $t$.

In the first sub-step (prediction) the position, $\vec{r}_\mathrm{pred}$,
and the velocity, $\vec{v}_\mathrm{pred}$, are estimated at time $t+dt$
by a truncated Taylor expansion using the position, velocity, acceleration
and jerk at time $t$
\begin{equation}
  \vec{r}_\mathrm{pred}(t+dt) =
  \vec{r}(t) + \vec{v}(t) dt  + \frac{1}{2}\vec{a}(t) dt^2
  + \frac{1}{6}\vec{j}(t) dt^3  
\end{equation}
and
\begin{equation}
  \vec{v}_\mathrm{pred}(t+dt) =
  \vec{v}(t) + \vec{a}(t) dt  + \frac{1}{2}\vec{j}(t) dt^2\,.
\end{equation}

In the second sub-step the  acceleration, $\vec{a}_\mathrm{pred}$, and jerk,
$\vec{j}_\mathrm{pred}$, are calculated at time $t+dt$ using
$\vec{r}_\mathrm{pred}$ and $\vec{v}_\mathrm{pred}$.

In the third step the higher order terms of the Taylor expansion at time $t$
\begin{equation}
  \vec{s}=\ddot{\vec{a}}=
  \frac{6(\vec{a}_\mathrm{pred}-\vec{a})}{dt^2}
  - \frac{2(\vec{j}_\mathrm{pred}+2\vec{j})}{dt}
\end{equation}
  (called snap) and
\begin{equation}
  \vec{c}=\dddot{\vec{a}}=
  \frac{12(\vec{a}-\vec{a}_\mathrm{pred})}{dt^3}
  + \frac{6(\vec{j}_\mathrm{pred}+\vec{j})}{dt^2}
\end{equation}
(called crackle) are calculated.

In the fourth step the higher terms are added to the predicted
position and velocity to obtain
corrected values at the time $t+dt$
\begin{equation}
  \vec{r}_\mathrm{corr}=\vec{r}(t+dt) = 
  \vec{r}_\mathrm{pred} + \frac{1}{24}\vec{s}(t) dt^4
  + \frac{1}{120}\vec{c}(t) dt^5  
\end{equation}
and
\begin{equation}
  \vec{v}_\mathrm{corr}=\vec{v}(t+dt) = 
  \vec{v}_\mathrm{pred} + \frac{1}{6}\vec{s}(t) dt^3
  + \frac{1}{24}\vec{c}(t) dt^4\;.  
\end{equation}
These corrected values are taken as new positions and velocities at time
$t+dt$.

\subsection{MONDian acceleration and jerk}
First, the MONDian equations of motion  (Eq.~(\ref{eq_mld_sub}))
have to be solved for the acceleration.
In Milgrom-law dynamics
the Newtonian and the MONDian acceleration vectors are collinear.
By introducing the unit acceleration vector $\vec{e}$ it follows
\begin{equation}
 \mu\left(\frac{|\vec{a}|}{a_0}\right)|a|\vec{e}
  =|g|\vec{e}\,.
\end{equation}
Dividing by $a_0$ and setting
\begin{equation}
  x = \frac{|\vec{a}|}{a_0}\;\;\;,\;\;\; y = \frac{|\vec{g}|}{a_0}
\end{equation}
\begin{equation}\label{eq_mu_x_y}
 \mu\left(x\right)x=y
\end{equation}
emerges.\\
1. case, $y=0$:
It follows $x=0$.\\
2. case, $y\neq 0$: The left hand side of Eq.~(\ref{eq_mu_x_y}) is an injective
mapping from $\mathbb{R}_{>0}$ onto $\mathbb{R}_{>0}$. Therefore, a unique
solution $y$ exists. The solution can be obtained numerically in the
general case. For some special forms of $\mu(x)$ Eq.~(\ref{eq_mu_x_y})
can be solved analytically for $x$.
Then, the MONDian acceleration vector can be calculated
by
\begin{equation}
  \vec{a} = \left|\vec{a}\right|\vec{e}
  =x a_0 \vec{e} =\frac{x}{y} y a_0 \vec{e}
  =\frac{x}{y}\left|\vec{g}\right|\vec{e}
  =\frac{x}{y}\vec{g}\,.
\end{equation}

The jerk, $\vec{j}$, of each particle can by calculated by
differentiation of the MONDian acceleration with respect to the physical
time.  Differentiation of Eq.~(\ref{eq_mlv}) leads to
\begin{equation}\label{eq_mu_prime}
  \mu^\prime\left(\frac{|\vec{a}|}{a_0}\right)
  \frac{\vec{a}\bullet\vec{j}}{a_0|\vec{a}|}\vec{a}
  +\mu\left(\frac{|\vec{a}|}{a_0}\right)\vec{j}
  = \dot{\vec{g}}\,.
\end{equation}
This equation contains the jerk as a vector and as a part of a scalar
product. By calculating the scalar product with $\vec{a}$ we
obtain an equation of $\vec{a}\bullet\vec{j}$ only which can
be solved for it,
\begin{equation}
  \vec{a}\bullet\vec{j} =
  \frac{\dot{\vec{g}} \bullet \vec{a}}{\mu^\prime(x)x+\mu(x)}\,,
\end{equation}
where the argument of the transition function is again $x=|\vec{a}|/a_0$.
Finally, the jerk can be obtained from Eq.~(\ref{eq_mu_prime})
after inserting the expression for $\vec{a}\bullet\vec{j}$
\begin{equation}
  \vec{j} = \left.\left(\dot{\vec{g}} -
  \frac{\mu^\prime(x)}{\mu^\prime(x)x+\mu(x)}\frac{\dot{\vec{g}}\bullet\vec{a}}{a_0 |\vec{a}|}
  \vec{a}\right)\middle/\mu(x)\right.\;.
\end{equation}

\subsection{Newtonian case}

The Newtonian case can be treated in the MONDian algorithm by setting
the critical acceleration, $a_0$, to a very small value. $a_0=10^{-20}$~Myr/pc$^2$ has been chosen
in this work for this case. Then for $|\vec{a}|\gg a_0$
$\mu(x)$ tends to 1, $\mu^\prime(x)$ tends to 0 and
the MONDian values converge against the Newtonian values,
$\vec{a}\,\to\,\vec{g}$ and
$\vec{j}\,\to\,\dot{\vec{g}}$.

\subsection{Transition function}
In this work we use the transition function 
\begin{equation}\label{eq_mu_here}
  \mu(x) = \frac{x}{\sqrt{1+x^2}}\,,
\end{equation}
which is commonly referred to as the standard interpolation function
\citep{famaey2012a},
with derivative
\begin{equation}
  \mu^\prime(x) = \frac{1}{\left(1+x^2\right)^{3/2}}\,.
\end{equation}
Eq.~(\ref{eq_mu_x_y}) can then be solved analytically,
\begin{equation}
  x = \sqrt{\frac{y^2+\sqrt{y^4+4y^2}}{2}}\;.
\end{equation}

\subsection{Newtonian acceleration and jerk with softening}
In the Newtonian context
the total (kinetic plus potential) energy of the individual stars
follows a distribution function. Stars with positive total energy are able
to escape from the star cluster and populate the tidal tails.
This region of the energy distribution function is continuously repopulated  
due to energy redistribution among the remaining members of the star cluster.
This  energy gain is mainly due to numerous
distant encounters between the stars.
However, very close encounters occur rarely in open star cluster
but pose numerical problems and require special algorithmic treatments
\citep{aarseth2003a}.
In order to avoid these laborious implementations
the singularity in the Newtonian potential, $U_{ji}$, between
two point masses, $m_j$ and  $m_i$,
is removed by
adding a softening parameter, $\varepsilon$,
to the distance of these particles \citep{aarseth1963a},
\begin{equation}
  U_{ji}=-G\frac{m_jm_i}{|\vec{r}_{ji}|^2+\varepsilon^2}\;.
\end{equation}
Here,
$\vec{r}_{ji}:=\vec{r}_j-\vec{r}_i$ is the separation vector between the $i$-th and
$j$-th
particle.
Then the total Newtonian acceleration, $\vec{g}_i$, of the $i$-th particle
is given by summing over all pairwise softened gravitational force contributions
\begin{equation}
  \vec{g}_i=G \sum_{\substack{j=0\\j\neq i}}^{j=n} \frac{m_j}{\left(\left|\vec{r}_{ji}\right|^2+\varepsilon^2\right)^{3/2}}\vec{r}_{ji}+\vec{g}_{\mathrm{ext},i}\,,
\end{equation}
where $\vec{g}_{\mathrm{ext},i}$ is the acceleration in the external field
at the position of the $i$-th particle, $\vec{r}_i$.
The corresponding softened Newtonian jerk is obtained by differentiation with
respect to time,
\begin{equation}
  \dot{\vec{g}}_i=G \sum_{\substack{j=0\\j\neq i}}^{j=n}
  \frac{m_j}{\left(\left|\vec{r}_{ji}\right|^2+\varepsilon^2\right)^{3/2}}
  \left(\vec{v}_{ji}-3\frac{\vec{r}_{ji}\bullet\vec{v}_{ji}}{\left|\vec{r}_{ji}\right|^2+\varepsilon^2}\vec{r}_{ji}\right)
  +\dot{\mathbf{g}}_\mathrm{ext}\,,
\end{equation}
where $\vec{v}_{ji}:=\vec{v}_j-\vec{v}_i$ is the relative velocity vector
between the particles $j$ and $i$.

Additional to the numerical reason the softening accomplishes a physical purpose. In case of a very close subsystem
the total particle accelerations are Newtonian and the $\mu$-factor is unity. Thus, an internally Newtonian subsystem
would follow a Newtonian orbit in the Galaxy rather than a MONDian orbit.
Thus, the softening avoids the Newtonisation of close subsystems and sets a limitation of this method.

\subsection{Galactic tidal field}

For the Galactic tidal field an entirely flat rotation curve
with a circular speed of $v_\mathrm{c}=225$\,km/s, the same as in
the related studies of open star clusters with asymmetric
tidal tails \citep{jerabkova2021a,pflamm-altenburg2023a}, is chosen.
The kinematical acceleration vector, $\vec{a}$,
as a function of the Galactocentric distance, $r$,
and in the case of a constant circular velocity, $v_\mathrm{c}$, in the $x$-$y$-plane
is
\begin{equation}\label{eq_a_flat}
  \vec{a} = -\frac{v_\mathrm{c}^2}{r}\,\vec{e}_r
  =-\frac{v_\mathrm{c}^2}{r^2}\,\vec{r}\;.
\end{equation}
The required Newtonian acceleration, $\vec{g}_\mathrm{ext}$, in order to keep
a particle in MONDian dynamics on a circular path is
\begin{equation}\label{eq_g_flat}
    \vec{g}_\mathrm{ext}=
  -\mu\left(\frac{v_\mathrm{c}^2/r}{a_0}\right)\frac{v_\mathrm{c}^2}{r^2}\vec{r}\,.
\end{equation}
For the interpolating function chosen here (Eq.~(\ref{eq_mu_here}))
the explicit acceleration is
\begin{equation}
  \vec{g}_\mathrm{ext} = -\frac{v_\mathrm{c}^4}{\sqrt{r^6a_0^2+r^4v_\mathrm{c}^4}}\vec{r}\,
\end{equation}
and the corresponding Newtonian external jerk is
\begin{equation}
  \dot{\vec{g}}_\mathrm{ext} =
    -\frac{v_\mathrm{c}^4}{\sqrt{r^6a_0^2+r^4v_\mathrm{c}^4}}\vec{v}
    +\frac{v_\mathrm{c}^4\;\vec{r}\bullet\vec{v}\;r^2\left(3a_0^2r^2+2v_\mathrm{c}^4\right)}{\left(r^6a_0^2+r^4v_\mathrm{c}^4\right)^{\frac{3}{2}}}\vec{r}\;.
\end{equation}
The Newtonian limit is again obtained by $a_0\to 0$ and the external Newtonian
acceleration converges against the right-hand side of Eq.~(\ref{eq_a_flat}).

\section{Numerical tests of MLD}\label{sec_num_tests}
In order to get an insight what are the effects of this formulation of MOND,
the dynamical behavior of special $N$-body systems are explored
in this section
before studying the formation and evolution of tidal tails in this dynamical
context.

\subsection{Isolated deep MOND binary}
\label{subsec_isolated_MLD_binary}
The first system considered is that of an isolated binary of non-equal mass
constituents in the deep MOND regime. A binary with masses
$m_1=2\,M_\odot$ and $m_2=0.2\,M_\odot$ is set up such that its semi-major axis
would be $a=1\,$pc and eccentricity $e=0.1$ in Newtonian dynamics with
an orbital period of $T =\sqrt{\frac{4\pi^2 a^3}{G(m_1+m_2)}}=63.15\,\rm Myr$.
The minimum distance would be $r_\mathrm{min} =  a (1-e)=0.9\,\rm pc$
and therefore the maximum internal acceleration of the less-massive
component is $g_\mathrm{max} = G m_1/r_\mathrm{min}^2=0.01\,\rm pc/Myr^2$,
which is two orders of magnitude smaller than $a_0$. The initial conditions
are $\mathrm{r}_1=(-0.0818182, 0, 0)\,\rm pc$ and
$\mathrm{v}_1=(0,-0.01,0)\,\rm pc/Myr$ for particle 1
and $\mathrm{r}_2=(0.818182,0,0)$ and $\mathrm{v}_2=(0,0.1,0)$ for particle 2.
The binary is integrated with no softening ($\varepsilon=0$).

\begin{figure*}
  \includegraphics[width=\columnwidth]{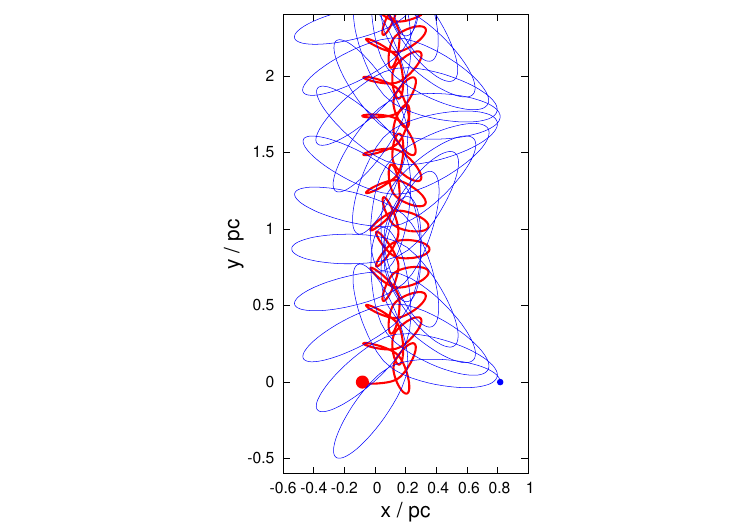}%
    \includegraphics[width=\columnwidth]{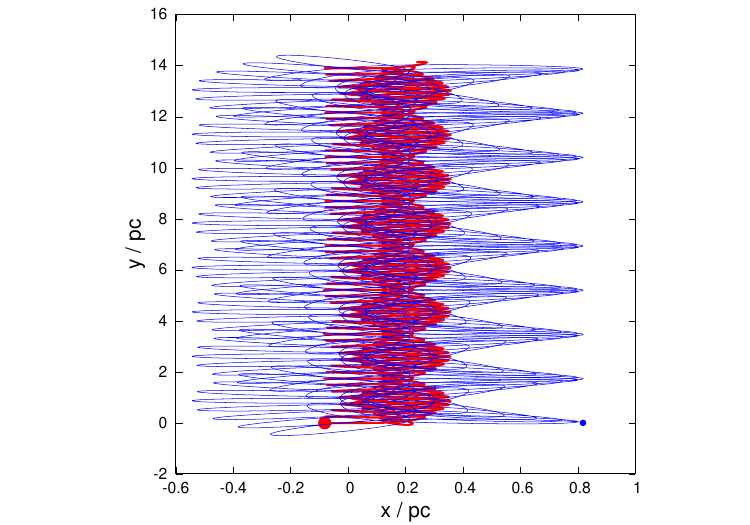}
    \caption{\label{fig_binary_orbit} Orbital evolution of
      a deep MOND MLD-binary: ({\it Left:}) The thick red curve shows
      the trajectory of the more massive particle with $m_1=2\,\rm M_\odot$,
      the thin blue curve shows the trajectory of the less massive particle
      with $m_1=0.2\,\rm M_\odot$. The large red filled
      circle indicates the initial position of the more massive particle, the
      small blue filled circle the initial position of the less massive
      particle. ({\it Right:}) Shown is the complete orbital evolution over
      a period of 1\,Gyr.
    }
\end{figure*}

In Newtonian dynamics stable and non-changing elliptical orbits are expected.
In contrast, the left panel of Fig.~\ref{fig_binary_orbit} shows a slightly
chaotic motion. The right panel
reveals a regular pattern on a larger time-scale. The MOND binary seems
not to be self-accelerated.
This can be understood by inspection of the equation of motion.
In the case that the system evolves in the deep MOND regime the
general MLD-equation can be approximated by 
\begin{equation}\label{eq_deep_mdl_two_body_1}
  \mathbf{a}_1 = \sqrt{Ga_0m_2} \frac{\mathbf{q_2}-\mathbf{q_1}}{|\mathbf{q_2}-\mathbf{q_1}|^2}
\end{equation}
\begin{equation}\label{eq_deep_mdl_two_body_2}
  \mathbf{a}_2 = -\sqrt{Ga_0m_1} \frac{\mathbf{q_2}-\mathbf{q_1}}{|\mathbf{q_2}-\mathbf{q_1}|^2}\,.
\end{equation}
It can be easily verified that these equations can be derived from the
Lagrangian
\begin{equation}
  L_\mathrm{MLD} = \frac{\sqrt{m_1}}{2}\dot{\vec{q}}_1^2
  + \frac{\sqrt{m_2}}{2}\dot{\vec{q}}_2^2
  -\sqrt{Ga_0m_1m_2}\ln\left(|\vec{q}_2-\vec{q}_1|\right)\,.
\end{equation}
Comparing with the Newtonian Lagrangian,
\begin{equation}
  L_\mathrm{N} = \frac{m_1}{2}\dot{\vec{q}}_1^2
  + \frac{m_2}{2}\dot{\vec{q}}_2^2
  +Gm_1m_2\frac{1}{|\vec{q}_2-\vec{q}_1|}\,,
\end{equation}
two changes can be seen. The Keplerian potential is replaced by a logarithmic
potential and the  masses of the particles in the kinetic potential turn into their square roots. 
Furthermore, inert and heavy mass are still equal.

\begin{figure*}
  \includegraphics[width=\columnwidth]{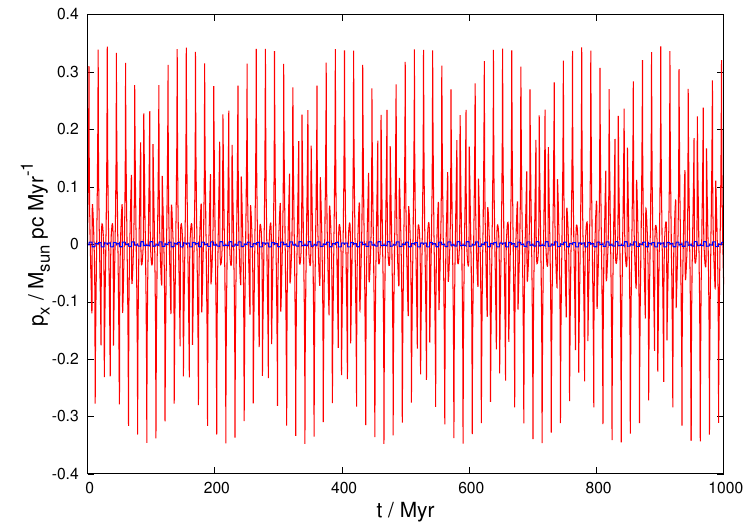}%
    \includegraphics[width=\columnwidth]{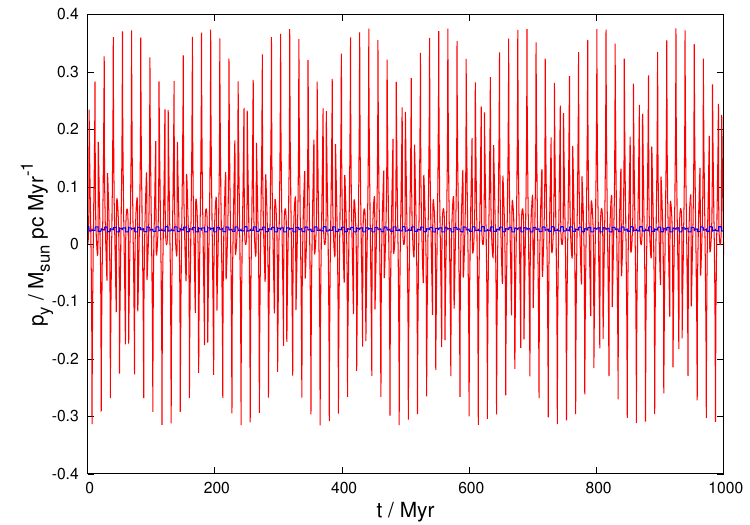}
    \caption{\label{fig_binary_momentum}
    Evolution of the linear momentum of
    a deep MOND MLD-binary: The slightly varying blue curve shows the
    MLD-linear momentum (Eq.~(\ref{eq_mld_lin_momentum}))
    as a function of time.
    The strongly oscillating red curve shows
    the time evolution of the corresponding Newtonian linear momentum,
    $\mathbf{p}_\mathrm{New}=m_1\dot{\mathbf{q}}_1+m_2\dot{\mathbf{q}}_2$.
    ({\it Left}): $x$-component of the linear momentum.
    ({\it Right}): $y$-component of the linear momentum.}
\end{figure*}

Additionally,
both Lagrangians share the same symmetries and are time-independent.
The translational symmetry leads to the conservation of the
MLD-momentum
\begin{equation}\label{eq_mld_lin_momentum}
  \mathbf{p}_\mathrm{MLD}=\sqrt{m_1}\dot{\mathbf{q}}_1+\sqrt{m_2}\dot{\mathbf{q}}_2 = \mathrm{const.}
\end{equation}
Figure~\ref{fig_binary_momentum} shows that the Newtonian linear momentum
does not increase continuously as the term self-acceleration might suggest,
but oscillates around a constant value. The MLD-linear momentum oscillates
very much weaker. It is not expected to stay constant as the binary evolves
in the deep MOND regime, but the MLD-equations of motion converge only
asymptotically against the
Eqs.~(\ref{eq_deep_mdl_two_body_1}) and (\ref{eq_deep_mdl_two_body_2})
for $\Delta q \to \infty$.

\begin{figure}
  \includegraphics[width=\columnwidth]{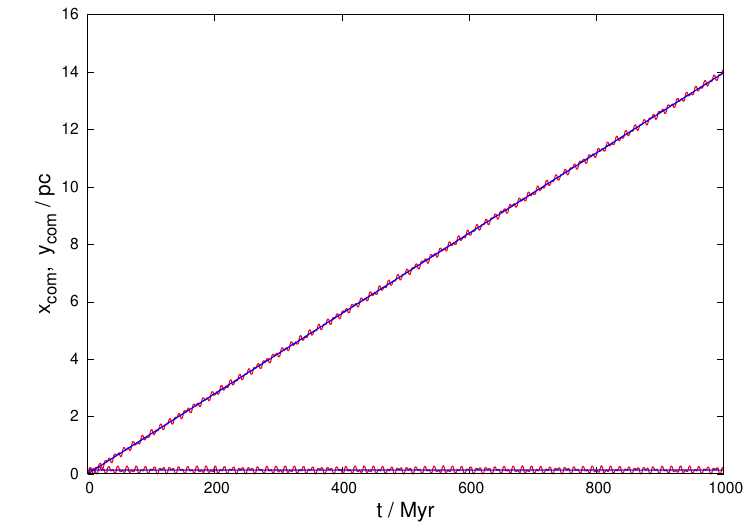}%
  \caption{\label{fig_binary_com} Centre of mass motions. 
    The straight blue lines refer to the MLD-centre of mass
    (Eq.~(\ref{eq_mld_com})) whereas the
    wobbling red curves show the Newtonian centre of mass,
    $\mathbf{R}_\mathrm{com,New}=\frac{{m_1}{\mathbf{q}}_1+{m_2}{\mathbf{q}}_2}{{m_1}+{m_2}}$.
    The $x$-component of both centers of mass runs horizontally,
    the $y$-component increases continuously.}
\end{figure}

The motion in space of the binary system can be obtained by considering
a transformation generated by
a boost of a velocity $\mathbf{v}$,
\begin{equation}
  \mathbf{q}_i^\prime(\varepsilon) = \mathbf{q}_i + \varepsilon\,\mathbf{v}t\,
  ,\,t^\prime = t\,.
\end{equation}
This leads to an invariant action and the quantity
\begin{equation}
  I = \left(\sqrt{m_2}\dot{\mathbf{q}}_1\bullet\mathbf{v}+\sqrt{m_2}\dot{\mathbf{q}}_2\bullet\mathbf{v}\right)t
  -
\left(\sqrt{m_1}\mathbf{v}\bullet\mathbf{q}_1+\sqrt{m_2}\mathbf{v}\bullet\mathbf{q}_2\right)
\end{equation}
is conserved for all boosts $\mathbf{v}$.
After devision by $(\sqrt{m_1}+\sqrt{m_2})$
a MLD-expression of a centre of mass  emerges
\begin{equation}\label{eq_mld_com}
  \mathbf{R}_\mathrm{com,MLD} =
  \frac{\sqrt{m_1}{\mathbf{q}}_1+\sqrt{m_2}{\mathbf{q}}_2}{\sqrt{m_1}+\sqrt{m_2}}\,,
\end{equation}
which moves with constant speed and constant direction
(Fig.~\ref{fig_binary_com}), whereas
the Newtonian centre of mass wobbles around the MLD-centre of mass.
 
\begin{figure}
  \includegraphics[width=\columnwidth]{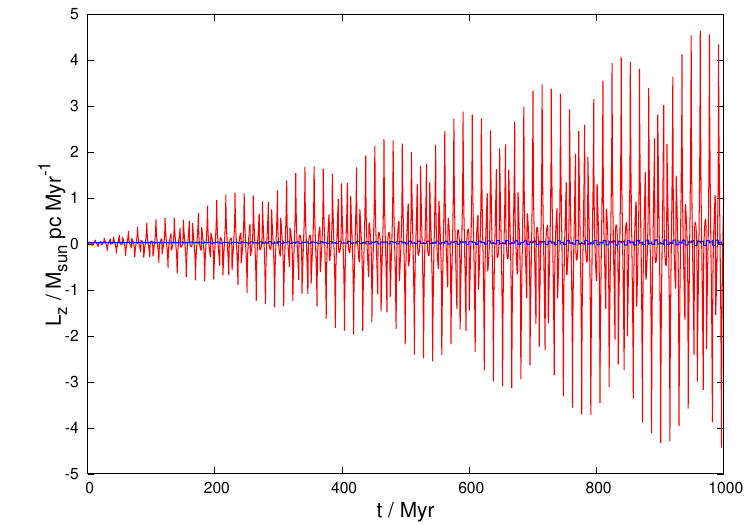}%
  \caption{\label{fig_binary_angular}
    Evolution of the angular momentum of
    a deep MOND MLD-binary: The slightly varying blue curve shows the
    MLD-agular momentum (Eq.~(\ref{eq_mld_ang_momentum}))
    as a function of time.
    The strongly oscillating red curve shows
    the time evolution of the corresponding Newtonian angular momentum,
    $\mathbf{L}_\mathrm{New}= m_1\mathbf{q}_1\times\dot{\mathbf{q}}_1
    +
    m_2\mathbf{q}_2\times\dot{\mathbf{q}}_2$.}
\end{figure}

Rotational symmetry leads to the conservation of the angular momentum
\begin{equation}\label{eq_mld_ang_momentum}
  \mathbf{L}_\mathrm{MLD}= \sqrt{m_1}\mathbf{q}_1\times\dot{\mathbf{q}}_1
  +
  \sqrt{m_2}\mathbf{q}_2\times\dot{\mathbf{q}}_2= \mathrm{const.}
\end{equation}
The evolution of the $z$-component of the Newtonian angular
momentum which oscillates increasingly with time is shown in
Fig.~\ref{fig_binary_angular}, whereas
the MLD angular momentum is almost conserved.

\begin{figure}
  \includegraphics[width=\columnwidth]{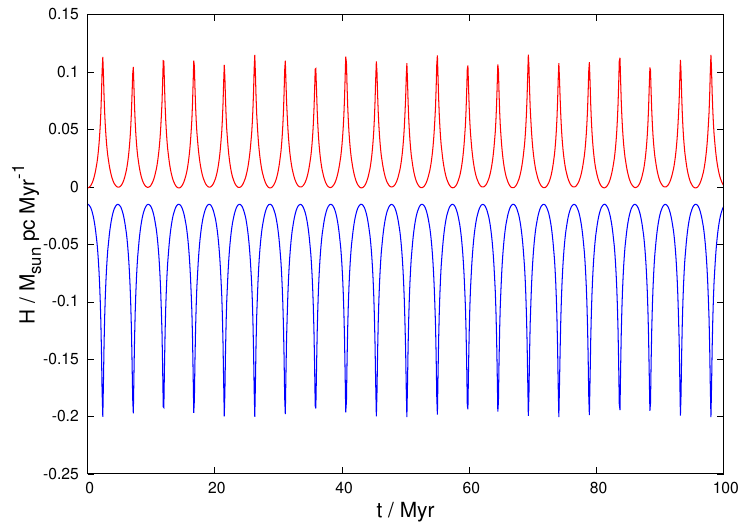}%
  \caption{\label{fig_binary_hamiltonian}
    Evolution of the Hamiltonian of
    a deep MOND MLD-binary: The blue curve shows the
    MLD-Hamiltonian (Eq.~(\ref{eq_mld_hamiltonian})) as a function of time.
    The red curve shows
    the time evolution of the corresponding Newtonian Hamiltonian,
    $H_\mathrm{New} = \frac{\mathbf{p}_1^2}{2m_1} +  \frac{\mathbf{p}_2^2}{2m_2}
  -G\;\frac{m_1m_2}{|\mathbf{q}_2-\mathbf{q}_1|}$.}
\end{figure}

The canonical momentum in MLD is 
\begin{equation}
  \mathbf{p}_i = \frac{\partial L_\mathrm{MLD}}{\partial \dot{\mathbf{q}}_i}
  = \sqrt{m_i}\dot{\mathbf{q}}_i
\end{equation}
and the associated Hamiltonian is  
\begin{equation}\label{eq_mld_hamiltonian}
  H_\mathrm{MLD} = \frac{\mathbf{p}_1^2}{2\sqrt{m_1}} +  \frac{\mathbf{p}_2^2}{2\sqrt{m_2}}
  -\sqrt{Ga_0m_1m_2}\;\ln\left(|\mathbf{q}_2-\mathbf{q}_1|\right)
  =\mathrm{const.}
\end{equation}
which is time-independent and is therefore conserved.
In this particular case both Hamiltonians oscillate (Fig.~\ref{fig_binary_hamiltonian}).
However, the time evolution show no secular evolution.

\subsection{Binaries in an external galactic field}
\begin{figure}    
  \includegraphics[width=\columnwidth]{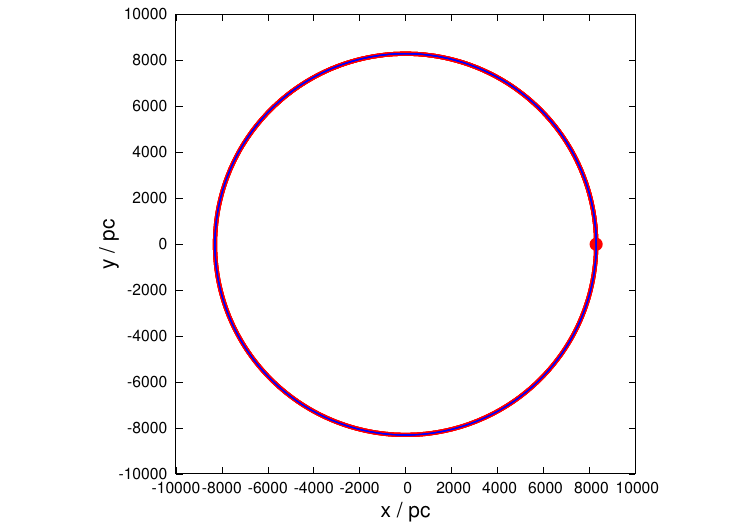}
  \caption{\label{fig_orbit_efe_wide_binary_mond} Binary in external field. The internally MONDian binary follows
    the MONDian Galactic orbit. The thick red curve shows the orbit of the $2\,M_\odot$-component, the thin blue curve
    shows the orbit of the less massive star. The filled red circle marks the initial position of the binary.}
\end{figure}

As a next test the internally MONDian binary from Sec.~\ref{subsec_isolated_MLD_binary}
is put on a circular orbit with radius $R=8300\,\rm pc$ in a flat rotation curve with
$v_\mathrm{c}=225\,\rm km/s = 230\,\rm pc/Myr$. The external acceleration is
$a_\mathrm{ext}=6.37\,\rm \rm pc/Myr^2$ (Eq.~\ref{eq_a_flat}) and the corresponding
Newtonian external acceleration is $g_\mathrm{ext} = 5.47\,\rm pc/Myr^2$ (Eq.~\ref{eq_g_flat}) and
therefore about two orders of magnitude larger than the internal acceleration.
Adding the kinematical circular velocity of 230\,pc/Myr to both components in Galactic tangential
direction leads to a circular Galactic motion of the internally MONDian binary (Fig.~\ref{fig_orbit_efe_wide_binary_mond})
with one Galactic revolution within 227\,Myr. The binary is integrated with no softening ($\varepsilon=0$).

The MLD-equation of motion of both binary components can be approximated by 
\begin{equation}\label{eq_mdl_two_body_1}
  \mu\left(\frac{a_\mathrm{ext}}{a_0}\right)\mathbf{a}_1 = \frac{G m_2}{|\mathbf{r}_2-\mathbf{r}_1|^2}(\mathbf{r}_2-\mathbf{r}_1) + \mathbf{g}_\mathrm{ext} 
\end{equation}
\begin{equation}\label{eq_mdl_two_body_2}
  \mu\left(\frac{a_\mathrm{ext}}{a_0}\right)\mathbf{a}_2 = -\frac{G m_1}{|\mathbf{r}_2-\mathbf{r}_1|^2}(\mathbf{r}_2-\mathbf{r}_1) + \mathbf{g}_\mathrm{ext} 
\end{equation}
in case if the external acceleration is very much larger than the internal acceleration, $\mathbf{a}_\mathrm{ext} >> \mathbf{a}_\mathrm{int}$.

For the acceleration of the Newtonian centre of mass follows
\begin{equation}
  \frac{m_1\mathbf{a}_1+m_2\mathbf{a}_2}{m_1+m_2} = \frac{\mathbf{g}_\mathrm{ext}}{\mu(a_\mathrm{ext}/a_0)} \,,
\end{equation}
and the radial acceleration is given by the scaled Newtonian acceleration.

\begin{figure}    
  \includegraphics[width=\columnwidth]{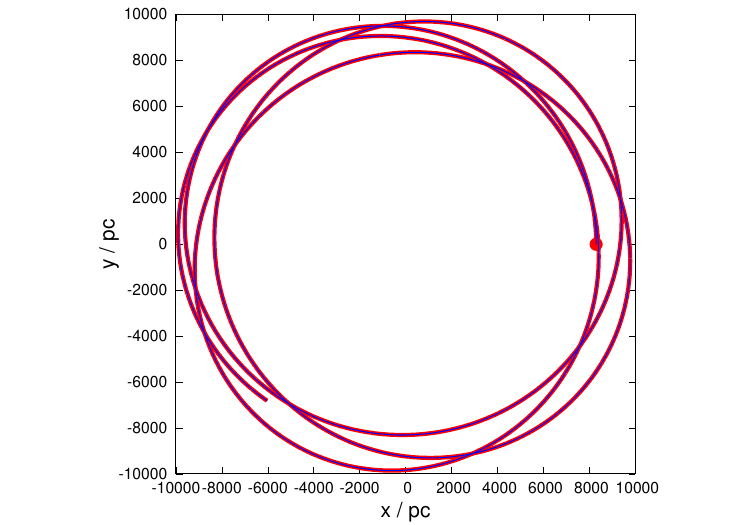}
  \caption{\label{fig_orbit_efe_wide_binary_newton} Binary in an external field. The internally Newtonian binary is set up 
    with a MONDian rotational velocity. The thick red curve shows the orbit of the $2\,M_\odot$-component, the thin blue curve
    shows the orbit of the less massive star. The filled red circle marks the initial position of the binary.}
\end{figure}

\begin{figure}    
  \includegraphics[width=\columnwidth]{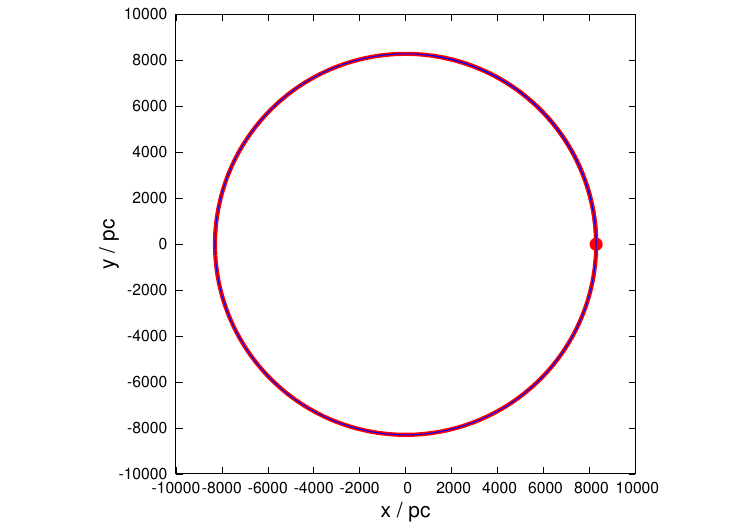}
  \caption{\label{fig_orbit_efe_wide_binary} Binary in an external field. The internally Newtonian binary is set up 
    with a Newtonian rotational velocity. The thick red curve shows the orbit of the $2\,M_\odot$-component, the thin blue curve
    shows the orbit of the less massive star. The filled red circle marks the initial position of the binary.}
\end{figure}

In the next step an internally Newtonian binary with the same components 
is set up with a semi major axis of $2\times 10^{-3}\,\rm pc$ with the same
tangential velocity of 230\,pc/Myr. The binary does not move on a circular orbit
as seen in Fig.~\ref{fig_orbit_efe_wide_binary_newton}. A circular MONDian velocity
of 230\,pc/Myr corresponds to a Newtonian circular velocity of 213\,pc.Myr at a Galactocentric distance
of 8300\,pc. Now, the binary moves on a circular orbit (Fig.~\ref{fig_orbit_efe_wide_binary}).
An internally Newtonian binary follows a Newtonian orbit.

\subsection{Isolated hierarchical triple}
An equal mass hierarchical triple is integrated in Newtonian dynamics and MLD, where the higher configuration is a MONDian binary with a
semi-major axis of 1\,pc and eccentricity $e=0$.
One component is a compact Newtonian binary with a
semi-major axis of 10$^{-3}$\,pc and eccentricity $e=0$. All three components have equal masses, $m_1=m_2=m_3=1\,M_\odot$. Thus, the internal
Newtonian binary is an equal-mass binary, the outer one is a non-equal-mass binary.
The initial conditions of the compact binary are
$\mathbf{r}_1 =(-0.333833,0,0)~\,\rm pc$, $\mathbf{v}_1 =(0,-1.53826,0)~\,\rm pc/Myr$ and
$\mathbf{r}_2 =(-0.332833,0,0)~\,\rm pc$, $\mathbf{v}_2 =(0,1.46083,0)~\,\rm pc/Myr$. The initial conditions of
the third body are $\mathbf{r}_3 =(0.666667,0,0)~\,\rm pc$, $\mathbf{v}_2 =(0,0.0774363,0)~\,\rm pc/Myr$.
The initial conditions are such that the Newtonian centre of mass is initially at rest at the origin.
The internal Newtonian acceleration of the compact binary is $a_\mathrm{in} = 4500\,\rm pc/Myr^2=1184\,a_0$. The acceleration
of the less massive component of the outer binary is $a_\mathrm{out} = 0.009\,\rm pc/Myr^2=0.0024\,a_0$.
The triple system is integrated for 100~Myr with no softening ($\varepsilon=0$).

\begin{figure}    
  \includegraphics[width=\columnwidth]{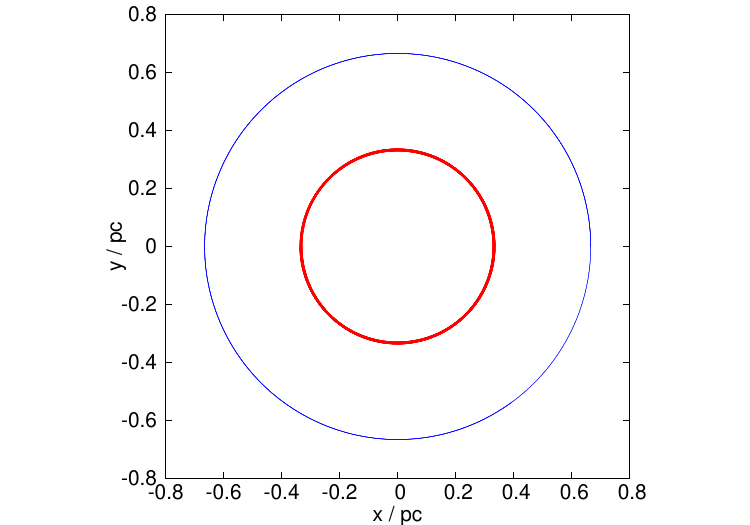}
  \caption{\label{fig_triple_new} Triple in Newtonian dynamics. The inner thick (red) circle shows the orbit of the inner more massive
  binary. The thin (blue) outer circle shows the orbit of the single star.}
\end{figure}

\begin{figure}    
  \includegraphics[width=\columnwidth]{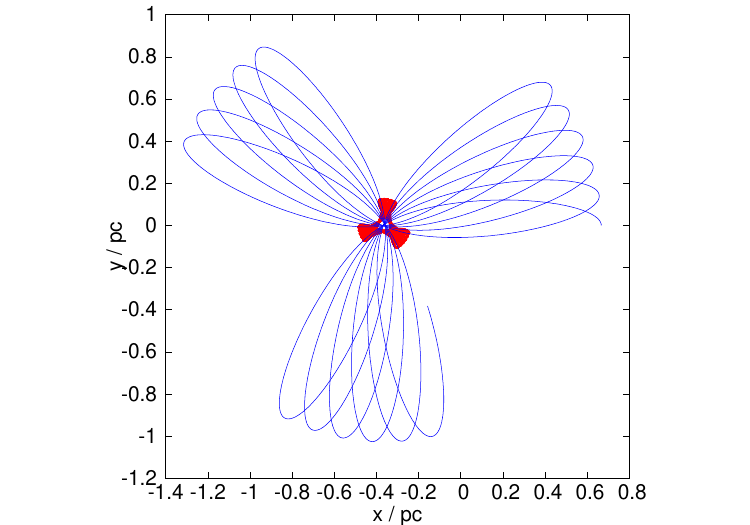}
  \caption{\label{fig_triple_mld} Triple in MLD. The inner thick (red) circle shows the orbit of the inner more massive
  binary. The thin (blue) outer circle shows the orbit of the single star.}
\end{figure}

Figure~\ref{fig_triple_new} shows the orbital evolution of the triple system integrated in Newtonian dynamics. As expected both
outer components move on circular orbits as it has no eccentricity. The centre of mass remains at rest. The evolution of the triple
system in MLD is shown in Fig.~\ref{fig_triple_mld}. The orbital configuration now precesses around the origin. No
net self-acceleration  is visible.

\subsection{Isolated Plummer sphere}
\begin{figure}    
  \includegraphics[width=\columnwidth]{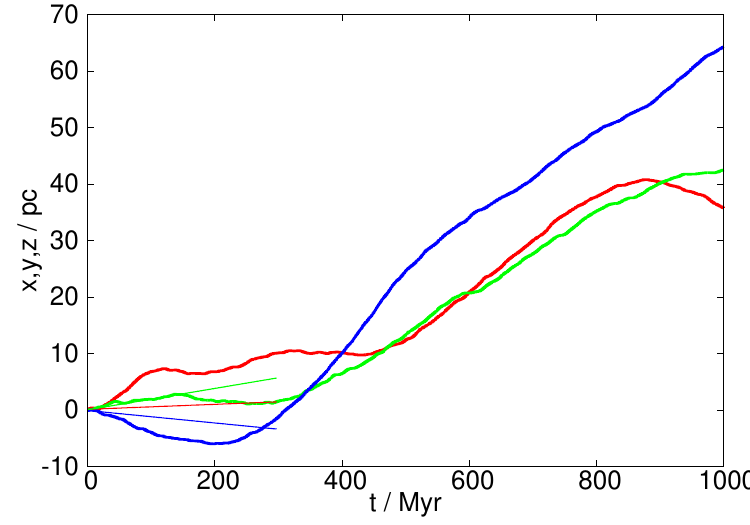}
  \caption{\label{fig_plummer_iso}Centre of mass evolution of an isolated MLD-Plummer sphere:
    The thick lines show the $x$-(red), $y$-(green) and $z$-(blue)component of the Newtonian centre of mass evolution with time of an isolated
    Plummer sphere in MLD. The thin lines show the evolution of the Newtonian centre of mass of a Plummer sphere in Newtonian
    dynamics with identical
    initial conditions as the MLD-Plummer sphere (with identical color-coding of the spatial components).}
\end{figure}

A Plummer sphere \citep{plummer1911a,aarseth1974a} is set up with $n=400$ particles with masses, $m_i = m_\mathrm{l}+(i-1) \Delta m$),
equally distributed from $m_\mathrm{l}=0.1~M_\odot$ and $m_\mathrm{u}=2~M_\odot$, with a mass difference of
$\Delta m = (m_\mathrm{u}-m_\mathrm{l})/(n-1)$. The total mass is $M_\mathrm{tot}=n(m_\mathrm{u}+m_\mathrm{l})/2=420\,M_\odot$.
Choosing a Plummer parameter of $b=3.1\,\rm pc$ the maximum internal acceleration is $0.076\,\rm pc/Myr^2 = 0.02\,a_0$.
Figure~\ref{fig_plummer_iso} shows the evolution with time of the three spatial components of the Newtonian centre of mass.
In lack of a known conserved quantity in MLD the MONDian
simulation is here compared with a Newtonian simulation
with identical initial conditions.

In the general $N$-body case it is currently not clear, if the MLD equations of motion can be derived from a
variational principle und therefore the search for conserved quantities is much more difficult. However, a generalized
expression of action equal reaction can be derived from Eq.~(\ref{eq_mld}) by multiplication with the mass of the considered
particle and by subsequent summation over all particles:
\begin{equation}
  \sum_{i=1}^{i=N}m_i\mu\left(\frac{|\mathbf{a}_i|}{a_0}\right)\mathbf{a}_i
  =G \sum_{\substack{j,i=0\\j\neq i}}^{j,i=N} \frac{m_im_j}{\left|\mathbf{r}_j-\mathbf{r}_i\right|^3}\left(\mathbf{r}_j-\mathbf{r}_i\right)
  =0\,.
\end{equation}
In the Newtonian limit the acceleration term can be expressed as the time derivative of the kinematical velocity
\begin{equation}
  0=
  \sum_{i=1}^{i=N}m_i\mathbf{a}_i=\sum_{i=1}^{i=N}m_i\dot{\mathbf{v}}_i = \frac{d}{dt}\sum_{i=1}^{i=N}m_i\mathbf{v}_i\,,
\end{equation}
and the total Newtonian linear momentum is conserved.

\subsection{Comparison with deep MOND two-body expressions}
\begin{figure}
  \includegraphics[width=\columnwidth]{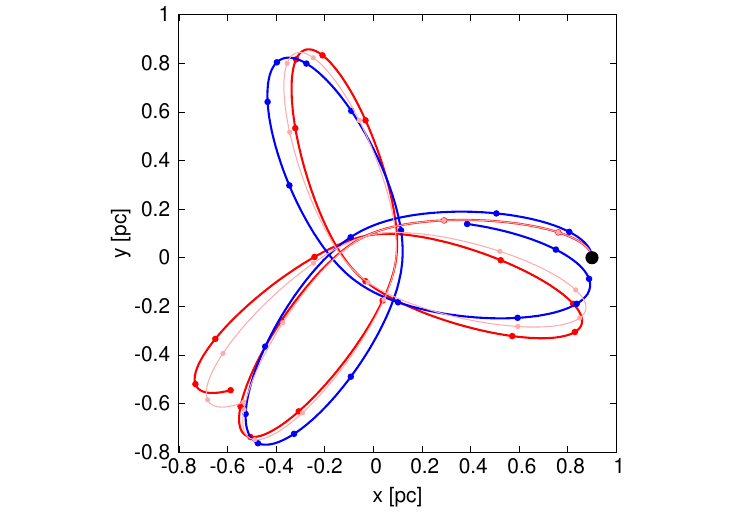}%
  \caption{\label{fig_deep_mond}Deep MOND binary. Shown is the relative
    motion over 20\,Myr of the isolated test binary with initial conditions
    (filled black circle) given at 
    the beginning of Sec.~\ref{subsec_isolated_MLD_binary} for three different
    sets of equations of motion: full MLD with transition function
    (red solid line), MLD  in the deep MOND limit
    (Eqs.~\ref{eq_deep_mdl_two_body_1} and \ref{eq_deep_mdl_two_body_2},
    light red line), and Milgrom's formulation
    (blue solid line,
    Eqs.~\ref{eq_deep_milgrom_two_body_1} and \ref{eq_deep_milgrom_two_body_2}).
    The small filled circles show the orbital positions in steps of 2\,Myr.
  }
\end{figure}

\citet[][ E.q 23]{milgrom2014a} concluded that any MOND field theory leads  
to an internal force of an isolated two-body system with masses $m_1$ and $m_2$
in the deep MOND limit of
\begin{equation}
  F = \frac{2\sqrt{a_0G}}{3 r_{21}}
  \left(
  (m_1+m_2)^{3/2}
  -m_1^{3/2}-m_2^{3/2}
  \right)\,,
\end{equation}
where $r_{21}$ is the interparticle distance.
Given that Newtonian momentum should be conserved in a MOND field theory
\citep{bekenstein1984a} $m_1\ddot{\mathbf{r}}_1=-m_2\ddot{\mathbf{r}}_2$
the equations of motion of both particles are
\begin{equation}
  \label{eq_deep_milgrom_two_body_1}
  \ddot{\mathbf{r}}_1=\frac{2\sqrt{a_0G}}{3m_1}
  \left(
  (m_1+m_2)^{3/2}
  -m_1^{3/2}-m_2^{3/2}
  \right)
  \frac{\mathbf{r}_2-\mathbf{r}_1}{|\mathbf{r}_2-\mathbf{r}_1|^2}
\end{equation}
and
\begin{equation}
  \label{eq_deep_milgrom_two_body_2}
  \ddot{\mathbf{r}}_2=-\frac{2\sqrt{a_0G}}{3m_2}
  \left(
  (m_1+m_2)^{3/2}
  -m_1^{3/2}-m_2^{3/2}
  \right)
  \frac{\mathbf{r}_2-\mathbf{r}_1}{|\mathbf{r}_2-\mathbf{r}_1|^2}
\end{equation}
The evolution of the relative orbit is shown in Fig.~\ref{fig_deep_mond}
in comparison to the MLD-equations for a duration of 20~Myr. In deep MOND the binary
following the \citet{milgrom2014a} equations of motion
has got a  slower motion in radial direction, but precesses slightly faster
than the MLD-binary.

\section{Tidal tail simulations of open star clusters}\label{sec_simulations}
In order to explore the difference in the dynamical formation and evolution
of tidal tails of open star clusters, low-mass star cluster models are set up
on a circular orbit in a Galactic field with a flat rotation curve
and are integrated with the algorithm presented in Sect.~\ref{sec_numerical_model}.

\subsection{Initial conditions}
The model star clusters contain 400 equal-mass point particles with a mass  of
0.5\,$M_\odot$ each, which are constant in time, i.e. stellar evolution is not included.
Thus, all models have a total mass of 200\,$M_\odot$.
The particles are set up in phase space as a Plummer model \citep{plummer1911a,aarseth1974a}
with a Plummer parameter of 3.1~pc. This corresponds to the properties
of the current Hyades star cluster \citep{roeser2019a}.

The initial position and velocity vectors of the star clusters are
chosen such that their
centre of mass orbits on a circular path with radius $r_0= 8300\,\rm pc$ in a
flat rotation curve with 225\,km/s, similar to the solar neighborhood
as used in related studies \citep{jerabkova2021a,pflamm-altenburg2023a}

\begin{figure}
  \includegraphics[width=\columnwidth]{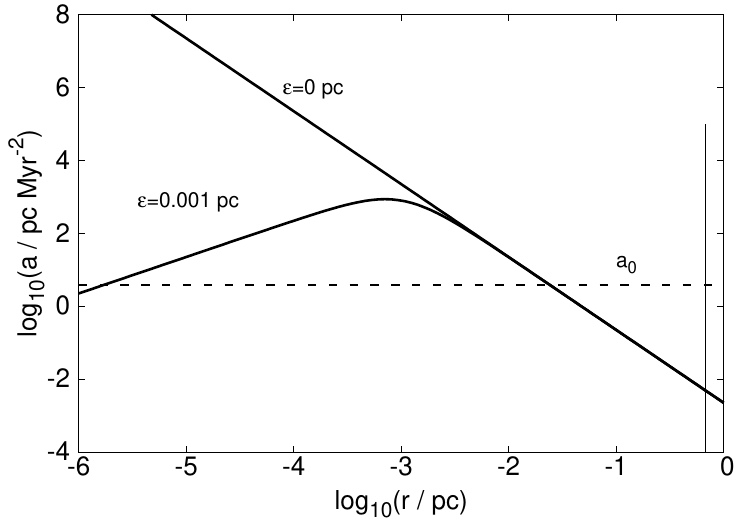}%
  \caption{\label{fig_smooth}Softening. Shown is the Newtonian acceleration field
    of a star with a mass of $0.5\,M_\odot$ in the case of no softening
    ($\varepsilon = 0\,\rm pc$) and softening with a parameter of
    $\varepsilon = 0.001\,\rm pc$. The dashed horizontal line marks
    the MONDian acceleration threshold, $a_0$. The vertical solid line indicates
    the mean central particle distance of $\approx 0.7$\,pc
    for a 400 particle Plummer sphere with Plummer
    parameter 3.1\,pc.
  }
\end{figure}

The smoothing parameter is set to $\varepsilon = 10^{-3}\,\rm pc$.
The difference between the softened and the unsoftened acceleration field of a
particle with a mass of 0.5\,$M_\odot$ is shown in Fig.~\ref{fig_smooth}. The
softened and the unsoftened
acceleration fields start to diverge from each other at an acceleration of about
100 times higher than the MONDian threshold. The central particle density of the
Plummer sphere is $n_0 = (3 N/4\pi b^3)$ corresponding to a mean central inter particle distance
of $\bar{d} \approx 0.7\,\rm pc$. Therefore, we expect that i) the driving long-distance
encounters responsible for energy redistribution and the consequent
evaporation of stars from the star cluster
and ii) the difference between MONDian and Newtonian dynamics below the acceleration
threshold, $a_0$,
are mostly unaffected by the gravitational softening.

In order to compare the MONDian and the Newtonian models,
the threshold value is a$_0=3.8$~Myr/pc$^2$
for the MONDian Models,
and $a_0=3.8\times 10^{-20}$~Myr/pc$^2$ for the Newtonian models.
In total 10 (5 MONDian and 5 Newtonian) models are calculated.
Two each models (1 Newtonian and 1 MONDian)
are created with the same random seed
and have identical initial conditions.

\subsection{Time evolution of the orbital snapshots}\label{sec_snap_shots}
\begin{figure*}
  \includegraphics[width=\columnwidth]{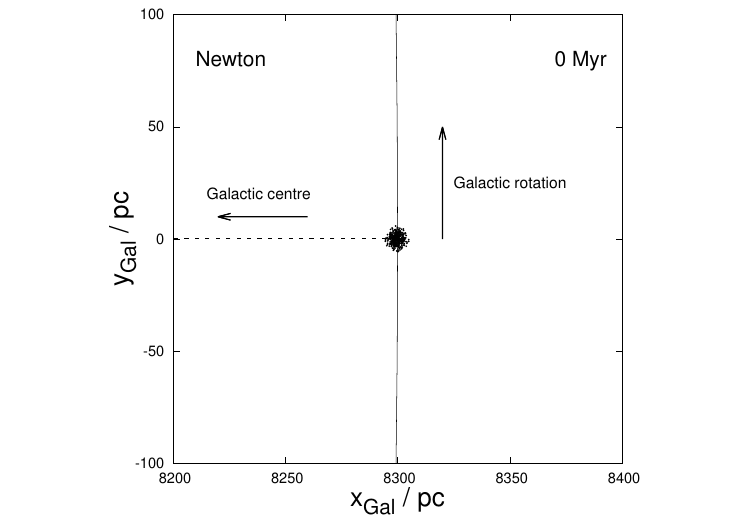}%
  \includegraphics[width=\columnwidth]{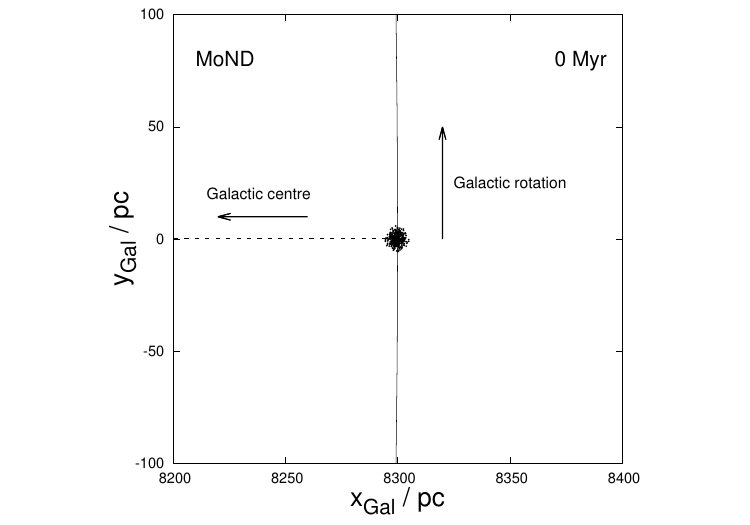}\\
  \caption{\label{fig_snap_0Myr}Orbital snapshots at 0~Myr.
    Star cluster evolution in Newtonian (left) and
    discrete Milgrom-law Dynamics (right). See Sect.~\ref{sec_snap_shots}
    for details.}
\end{figure*}

\begin{figure*}
  \includegraphics[width=\columnwidth]{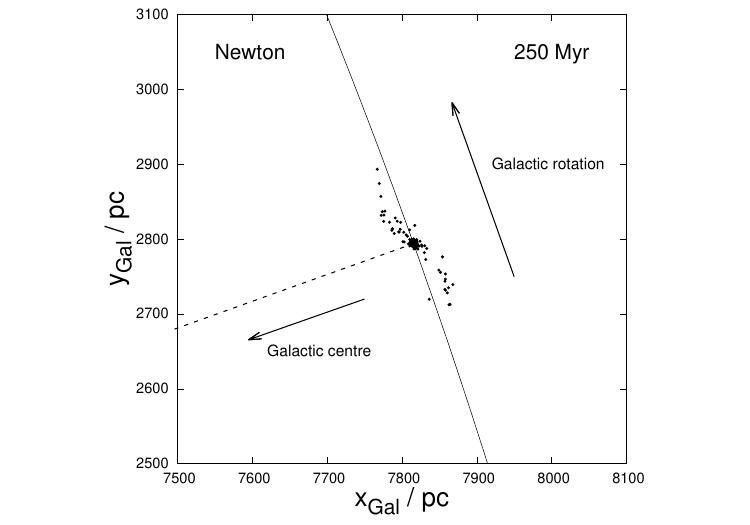}
  \includegraphics[width=\columnwidth]{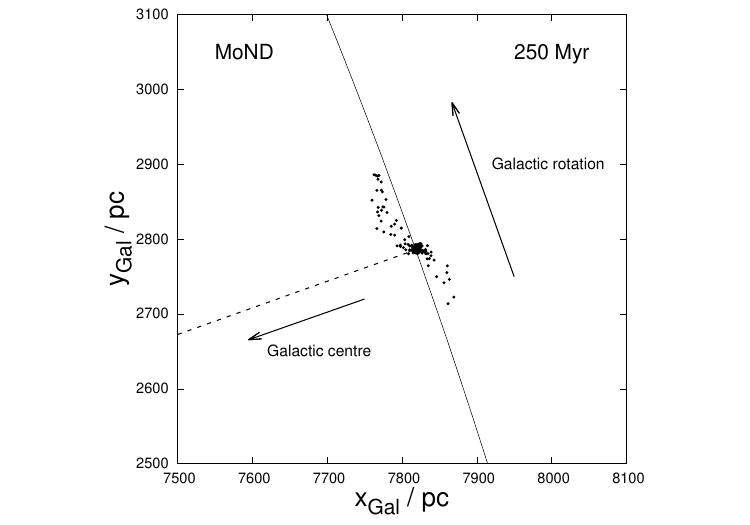}%\\
  \caption{\label{fig_snap_250Myr} Same as in Fig.~\ref{fig_snap_0Myr}
    but at 250~Myr.}
\end{figure*}

\begin{figure*}
  \includegraphics[width=\columnwidth]{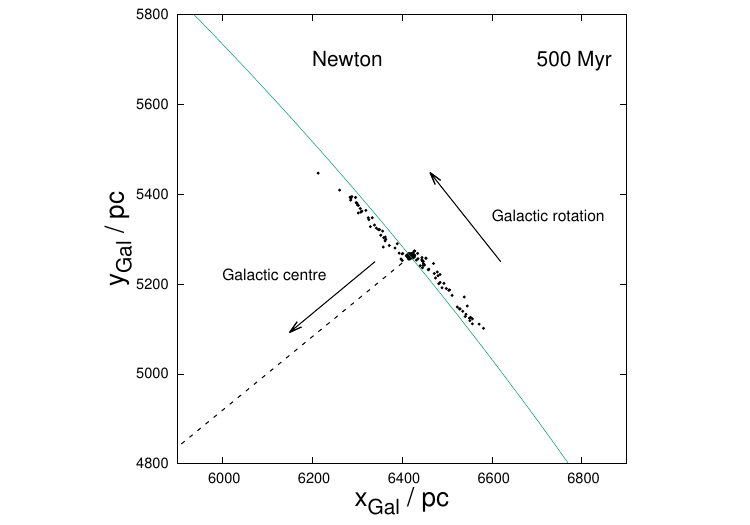}%
  \includegraphics[width=\columnwidth]{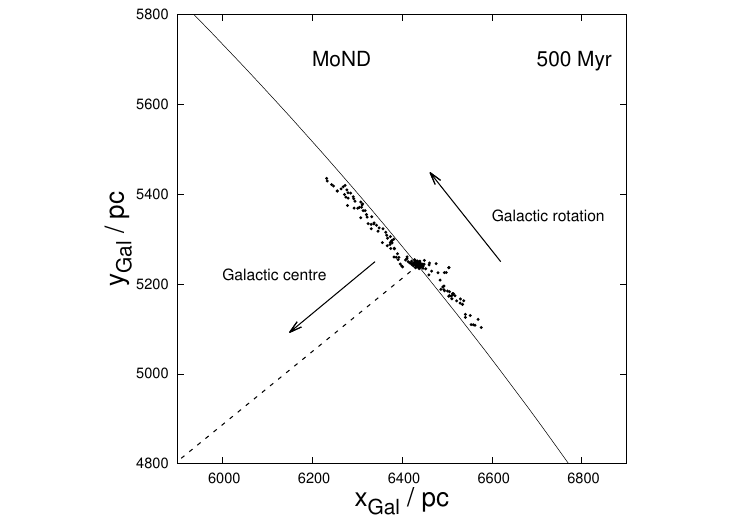}%
\caption{\label{fig_snap_500Myr} Same as in Fig.~\ref{fig_snap_0Myr}
  but at 500~Myr.}
\end{figure*}

\begin{figure*}
  \includegraphics[width=\columnwidth]{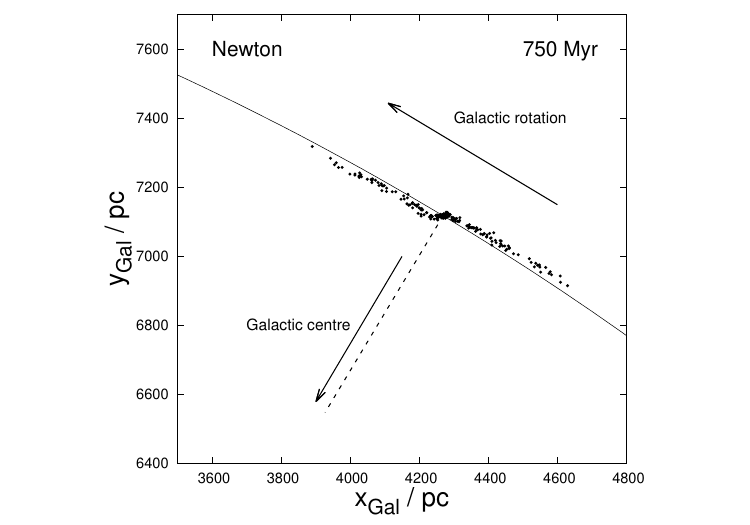}%
  \includegraphics[width=\columnwidth]{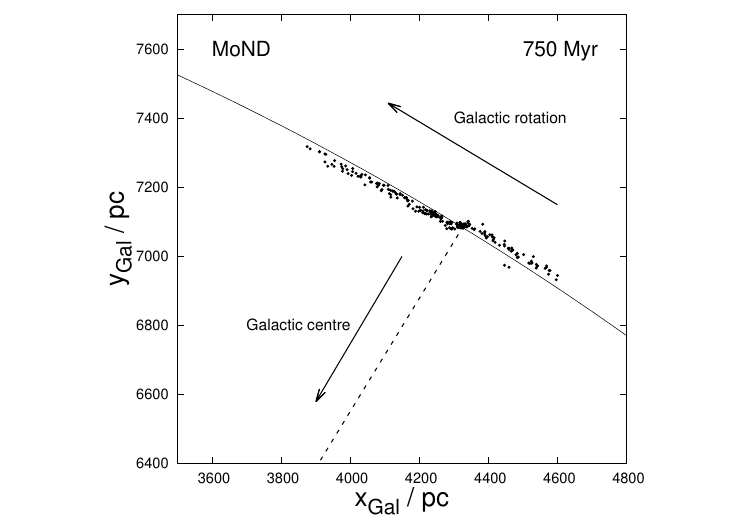}%
  \caption{\label{fig_snap_750Myr} Same as in Fig.~\ref{fig_snap_0Myr}
    but at 750~Myr.}
\end{figure*}

\begin{figure*}
  \includegraphics[width=\columnwidth]{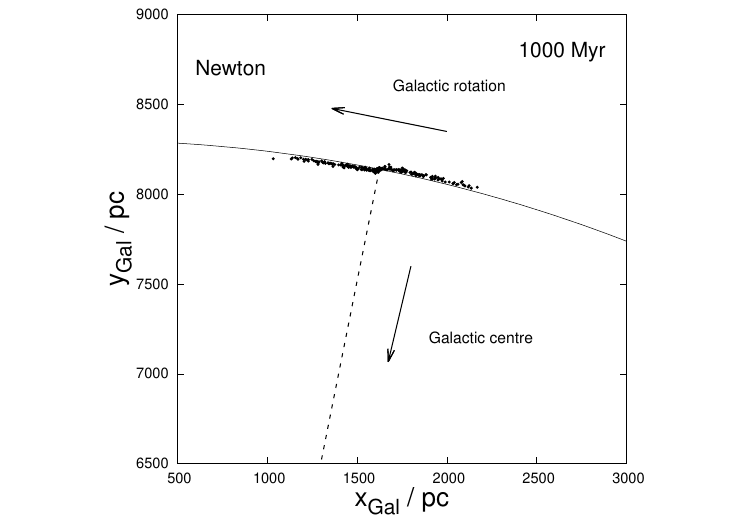}%
  \includegraphics[width=\columnwidth]{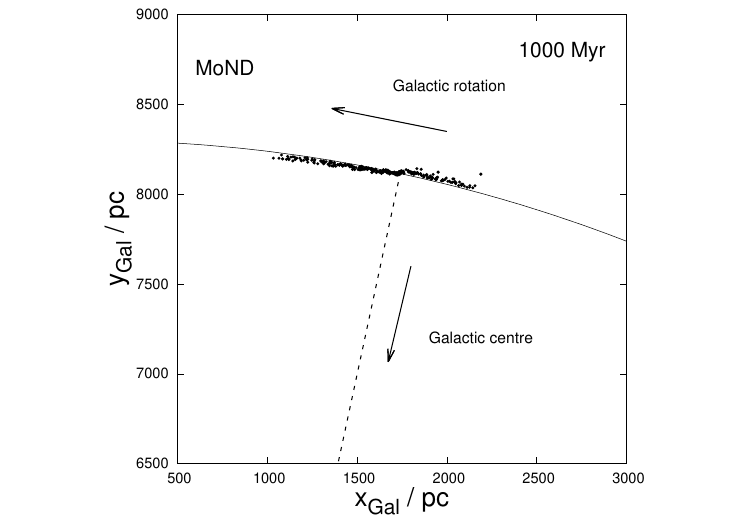}%
  \caption{\label{fig_snap_1000Myr} Same as in Fig.~\ref{fig_snap_0Myr}
    but at 1000~Myr.}
\end{figure*}

The time evolution  of one pair of star clusters in Newtonian and ML-dynamics
is displayed in Fig.~\ref{fig_snap_0Myr} at 0~Myr, in Fig.~\ref{fig_snap_250Myr}
after 250~Myr, in Fig.~\ref{fig_snap_500Myr} after 500~Myr,
in Fig.~\ref{fig_snap_750Myr} after 750~Myr, and in Fig.~\ref{fig_snap_1000Myr}
after 1000~Myr. In order to compare the relative positions of the star clusters
in Newtonian and ML-dynamics, the arrows indicating the direction of the
Galactic rotation and the direction to the Galactic centre and
the corresponding labels are located at the same position.
The short dashed line indicates the path from the Galactic centre to the 
density centre of the star cluster 
and the solid curved line shows the circular orbit.


The star clusters orbit anti-clockwise. The inner arm is the leading arm, the
outer arm is the trailing arm. In the Newtonian case both tidal arms are
approximately equally populated. In the MLD case the tidal tails are
populated asymmetrically. The leading tidal arm contains continuously
more members than the trailing arm.

Simultaneously to the asymmetric population of the tidal arms in ML-dynamics,
the star cluster follows the Newtonian star cluster. This can be seen in the larger
separation between the arrow pointing towards the Galactic centre and
the connecting (dashed)  line  between the Galactic centre and the density centre
of the star cluster in the MLD case than in the Newtonian case.

This might be due to some kind of local conservation of linear momentum.
As more stars end up in the leading tail, more momentum is carried away from
the star cluster into the moving direction of the star cluster. Thus, the star
cluster is expected to get a small recoil. However, as pointed out in
Sect.~\ref{sec_eqm} a concept like linear momentum conservation as in Newtonian
dynamics does not exist in Milgrom-law dynamics. Further theoretical
work is required in order to explore the conservation of dynamical
quantities in MLD.

\subsection{Analyzing the tidal tails}
The criterion whether or not a star is considered to be a member of the star
cluster or of the leading/trailing tidal arm is the same as in
\citet{pflamm-altenburg2023a}. If the distance of a star to the centre of the
star cluster is less than a cut of radius (here 10~pc) then the star is considered
to be a member of the star cluster. If this distance is larger than 10~pc then
the star is considered to be a member of one tidal arm. If the angle between
the distance vector of the particular star to the star cluster centre
and the velocity vector of the star cluster is less than 90$^\circ$
then the star is considered to be a member of the leading tidal arm.
If this angle is larger than 90$^\circ$ the star is assigned to
the trailing arm.

In \citet{pflamm-altenburg2023a} the stochastic asymmetry of tidal tails in Newtonian
dynamics has been quantified using test particle calculations in an analytic
Plummer potential orbiting the Galactic centre. The origin of the Plummer
potential has been used as the cluster centre.
In this work the cluster centre has to be determined by the positions of all
particles following standard procedures. 
A local density is assigned to each particle using the 6-th nearest neighbour
method \citep{casertano1985a}. In the next step
the density centre is generally calculated by the weighted sum of all local
particle densities \citep{aarseth2003a}. This method requires, that the
systems has nearly spherical symmetry. But due to the asymmetric tidal tails
this requirement is not met. Therefore, we first determine the position
of maximum density, which is expected to be located close to the centre of the
star cluster. Then, only those particles are considered for the calculation
of the density centre which lie within a sphere with 10~pc radius where the
position of maximum density is at the centre.

\subsection{Time evolution of the asymmetry of the tidal tails}
\begin{figure}
  \includegraphics[width=\columnwidth]{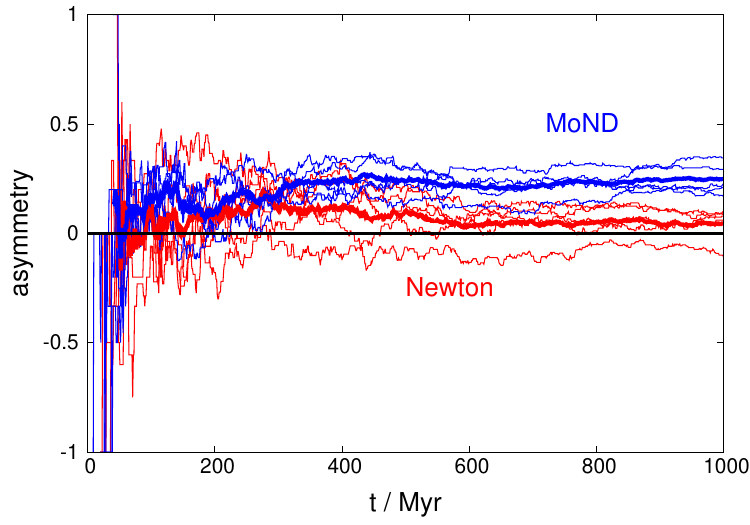}%
  \caption{\label{fig_asymmetry} Asymmetry of tidal arms. Thin lines show the
    evolution of the asymmetry of the 10 individual simulations, 5 Newtonian
    (red) and 5 MONDian (blue) simulations. The thick lines show the
    arithmetic mean values.}
\end{figure}

The asymmetry of the tidal arms at time $t$ is calculated by
\begin{equation}
  \epsilon = \frac{n_\mathrm{l}-n_\mathrm{t}}{n_\mathrm{l}+n_\mathrm{t}}\,,
\end{equation}
where $n_\mathrm{l}$ is the number of stars in the leading arm and
$n_\mathrm{t}$ is the number of stars in the trailing arm at time $t$
\citep{pflamm-altenburg2023a}. Thus, for $\epsilon > 0$ the leading arm contains
more members than the trailing arm. Fig.~\ref{fig_asymmetry} shows the evolution
of the asymmetry of all 10 simulations. It can be seen that in both dynamical
contexts the mean asymmetry is positive, the leading arm contains more members than
the trailing arm. But in MONDian dynamics the mean asymmetry varies between
0.2 to 0.3, whereas in Newtonian dynamics the mean asymmetry decreases continuously and slowly below
0.1. 

Comparison of the asymmetry
of the simulated star clusters with observed star clusters requires
that all tidal tail stars of a star cluster can be identified among the
field stars. This becomes more difficult with increasing distance of a tidal
tail star to the parent star cluster. Therefore, \citet{kroupa2022a}
considered only stars within a 50--200\,pc distance to the star cluster
and introduced the $q$-parameter
\begin{equation}
  q = \frac{N_\mathrm{l,50-200\,pc}}{N_\mathrm{t,50-200\,pc}}\,,
\end{equation}
where $N_\mathrm{l,50-200\,pc}$ is the number of stars in the leading arm in the
considered distance range and $N_\mathrm{t,50-200\,pc}$ the respective
number of stars in the trailing arm. For a $q$-parameter $q>1$ the leading arm
contains more members than the trailing arm in the respective distance range.

\begin{figure}
  \includegraphics[width=\columnwidth]{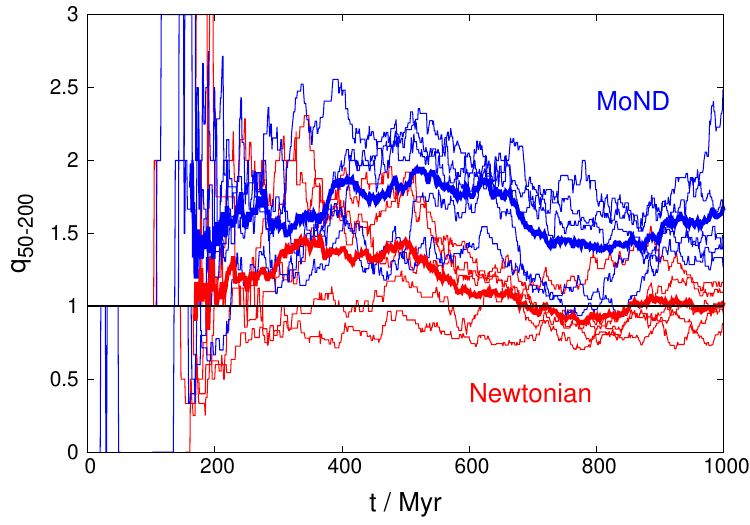}%
  \caption{\label{fig_q-parameter} $q$-parameter.  Thin lines show the
    evolution of the  $q$-parameter of the 10 individual simulations, 5 Newtonian
    (red) and 5 MONDian (blue) simulations. The thick lines show the
    arithmetic mean values.}
\end{figure}

The evolution of the $q$-parameter can be seen in Fig.~\ref{fig_q-parameter}. 
In MOND the $q$-parameter varies between 1.5 and 2. The leading arm contains
more stars than the trailing arm. In contrast,
in Newtonian dynamics the $q$-parameter finally stays constant around 1 and
the tidal tails show no asymmetry.

\subsection{Evaporation rate}
\begin{figure}
  \includegraphics[width=\columnwidth]{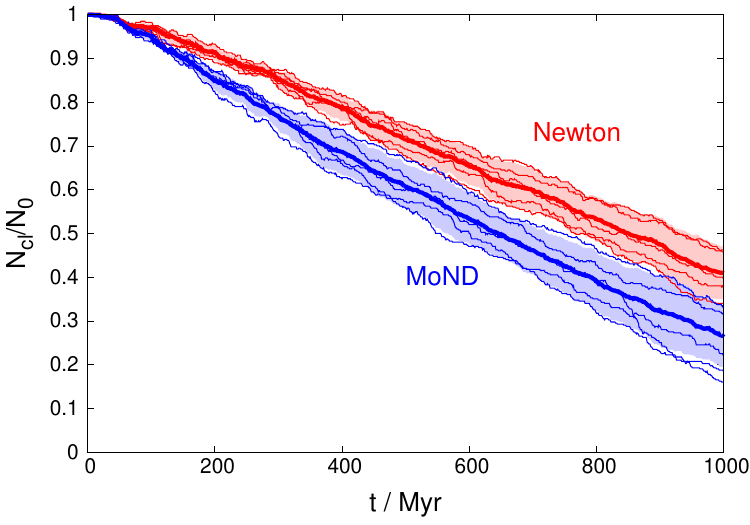}%
  \caption{\label{fig_dissolution} Evolution of the cluster member number
    fraction.
    Each of the 10 models  (5 Newtonian (red) and 5 MONDian (blue))
    models are shown by thin solid lines. The evolution of the mean values
    are given by thick solid lines. The light-blue and the light-red areas indicate
    the 1\,$\sigma$ region around the mean value.}
\end{figure}

The number of stars within the constant evaporation radius of 10\,pc
decrease differently fast
in both dynamical theories as shown in Fig.~\ref{fig_dissolution}.
In the MOND model the mean ratio of the current
number to the initial number 
of cluster members after 1~Gyr is 0.26 with a \,1~$\sigma$ variance of
0.07. Thus, 74\% of the initial stars have already evaporated from the cluster.
For 400 initial stars the evaporation rate in the MOND case is
0.296\,stars/Myr or one star every 3.4\,Myr.

The star cluster evaporates slower in the pure Newtonian case.
After 1\,Gyr of evolution the mean remaining cluster fraction is
0.41 with a 1~\,$\sigma$ variance of 0.06. Here, only 59\% of all stars
have evaporated from the star cluster within 1\,Gyr.
The evaporation rate is 0.236\,stars/Myr or one star every 4.2\,Myr. 
The evaporation rate in the MONDian case
is about 25\% larger than in the Newtonian case.

This can be understood by the increased effective internal
gravitational constant. The external kinematical acceleration
of a flat rotation curve with 225\,km/s = 230\,pc/Myr at a distance
of $8300\,\rm pc$ from the galaxy centre is
$a_\mathrm{ext}= 6.4\,\rm pc/Myr$. 

For a mass of $\approx 200\,M_\odot$ and
Plummer parameter $b=3.1\,\rm pc$ the internal maximal acceleration
is 0.036~pc/Myr$^2$.
Therefore, the internal dynamics is to first order
Newtonian with a gravitational constant increased
by a factor of $\mu(a_\mathrm{ext}/a_0)^{-1}= 1.16$.
From \citet[][Eq.~5]{baumgardt2003a} it is derived that
the dissolution time scale, $T_\mathrm{diss}$,
of a star cluster in Newtonian dynamics scales inversely with
the square root of the gravitational constant, $T_\mathrm{diss}\propto G^{-1/2}$.
The ratio of the dissolution time scales in the Newtonian case, $T_\mathrm{diss}$,
and in MLD in case of the external field effect, $T_\mathrm{diss,EFE}$,
is
\begin{equation}
  \frac{T_\mathrm{diss,EFE}}{T_\mathrm{diss}} = \sqrt{\frac{G}{G_\mathrm{EFE}}}
  =\sqrt{\mu(a_\mathrm{ext}/a_0)}
  =0.93\,.
\end{equation}
The ratio of the numerical dissolution time scale of the models here
is approximately
\begin{equation}
  \frac{T_\mathrm{diss,EFE}}{T_\mathrm{diss}} = \frac{\dot{N}}{\dot{N}_\mathrm{MLD}}
  = \frac{0.236}{0.296}= 0.8\,,
\end{equation}
and is about 10\% smaller than expected from the EFE approximation.

\section{Conclusions}

Direct $N$-body simulations of star clusters are performed
in a MONDian dynamical context.
For this Milgrom's law has been postulated to be valid for arbitrary
$N$-body systems.
In comparison to Newton models two main differences emerge:
i) The tidal arms are asymmetrically populated. The leading tidal arm contains
significantly more stars than the trailing arm. ii) The star cluster evaporates
stars or dissolves significantly faster.

The fact that asymmetric tidal tails arise in two different types
of MONDian dynamics, that is in
QUMOND \citep{thomas2018a,kroupa2022a} and Milgrom-law dynamics (this work),
leads to the conclusion that this effect is a property of the general
MONDian dynamical concept, rather than a result of the detailed realization.

However, the application of Milgrom's law to the dynamics of discrete N-body
systems
is limited to systems which are internally not Newtonian if embedded in
an external MONDian field.
This is here
artificially achieved by softening the two-body force.
Generalized Newtonian equations of motion should already contain
internally the non-Newtonisation
of centre of mass motions of compact subsystems, if constructible at all,
which needs to be explored.

However, in order to simulate the dynamical evolution of non-relaxed discrete systems
like open stars clusters or wide binaries the development of
collisional direct $N$-body codes are required as collision-less methods 
are not suitable for modeling these kind of systems.

This can be done by deriving mathematically consistent equations of motions for point-mass systems
from MOND field theories like AQUAL and QUMOND. But due to the non-linearity this is very difficult
to achieve. The other way would be, to extend the Newtonian equations of motions in such a way that
all necessary conditions of MOND-type theories are fulfilled.
Both strategies establish a complete new field of research.

Finally, because open star clusters are nearby systems
they are very well accessible
via astrometric observations.
and are ideal test objects
to discriminate between the validity of Newtonian or MONDian dynamics
on parsec scales.

\begin{acknowledgements}
  JPA acknowledges permanent hospitality by the Helmholtz-Institut
  f\"ur Strahlen-
  und Kernphysik.
\end{acknowledgements}

\bibliography{paperref}{}
\bibliographystyle{aa}

\end{document}